\documentclass[pra,twocolumn,preprintnumbers,amsmath,amssymb,nofootinbib]{revtex4}
\usepackage[T1]{fontenc}
\usepackage[latin1]{inputenc}
\usepackage{amsmath,amsfonts, graphicx,epsfig, exscale, relsize,pstricks,color}
\usepackage{slashed}
\usepackage{bbm,bm}
\usepackage{hyperref}

\newcommand{\pt}{\mathcal{T}}
\newcommand{\nL}{{n_{\text{L}}}}
\newcommand{\bgam}{\boldsymbol{\gamma}}

\newcommand{\Eqref}[1]{Eq.~\eqref{#1}}

\newcommand{\const}{c}

\begin{document}

\title{Interplay between geometry and temperature for inclined Casimir plates}
\author{Alexej Weber${}^1$ and Holger Gies${}^{2}$}

\affiliation{
\mbox{\it ${}^1$Institut f{\"u}r Theoretische Physik, Universit{\"a}t Heidelberg,
Philosophenweg 16, D-69120 Heidelberg, Germany} 
\mbox{\it ${}^2$Theoretisch-Physikalisches Institut, Friedrich-Schiller-Universit{\"a}t Jena,
Max-Wien-Platz 1, D-07743 Jena, Germany}
\mbox{\it E-mail: {a.weber@thphys.uni-heidelberg.de, holger.gies@uni-jena.de}}
}

\begin{abstract}
  We provide further evidence for the nontrivial interplay between geometry
  and temperature in the Casimir effect.  We investigate the temperature
  dependence of the Casimir force between an inclined semi-infinite plate
  above an infinite plate in $D$ dimensions using the worldline
  formalism. Whereas the high-temperature behavior is always found to be
  linear in $T$ in accordance with dimensional-reduction arguments, different
  power-law behaviors at small temperatures emerge. Unlike the case of
  infinite parallel plates, which shows the well-known $T^D$ behavior of the
  force, we find a $T^{D-1}$ behavior for inclined plates, and a $\sim
  T^{D-0.3}$ behavior for the {edge effect in the} limit where the plates
  become parallel. {The strongest temperature dependence $\sim T^{D-2}$ occurs for the
  Casimir torque of inclined plates.} Numerical as well as
    analytical worldline results are presented.
\end{abstract}

\maketitle

\section{Introduction}

The Casimir effect \cite{Casimir:dh} is not only a field witnessing rapid
experimental as well as theoretical progress, it also continues to offer
surprising new features. The Casimir effect derives its fascination from the
fact that it originates from quantum fluctuations of the radiation field or of
the charge distribution on the mesoscopic or macroscopic test
bodies. Moreover, it inspires many branches of physics, ranging from
mathematical to applied physics, see
\cite{Bordag:2001qi,Milton:2001yy} for reviews and \cite{Lamoreaux:1996wh} for
experimental verifications.

A distinctive feature of Casimir forces between test bodies is the dependence
on the geometry, i.e., the shape and orientation of these bodies.  For a
comparison between theory and a real Casimir experiment, a number of
properties such as finite conductivity, surface roughness and finite
temperature have to be taken into account in addition. Generically, these
latter corrections do not factorize but take influence on one another. For
instance, the interplay between dielectric material properties and finite
temperature \cite{Sernelius} is still {a subject of intense theoretical
  investigations} and has created a long-standing controversy
\cite{Mostepanenko:2005qh,Brevik:2006jw,Bimonte:2009nf,Ingold}. Also the role of
electrostatic patch potentials has been suggested as a potentially problematic
issue \cite{Speake:2003,Onofrio:2008}, which has become a matter of severe
debate \cite{Decca:2008,Onofrio:2009}.

The present article is not meant to resolve these controversies. On the
contrary, our work intends to draw attention to another highly nontrivial
interplay which on the one hand needs to be accounted for when comparing theory
and a real experiment and on the other hand is another characteristic feature
of the Casimir effect: the interplay between geometry and temperature. As
first conjectured by Jaffe and Scardicchio \cite{Scardicchio:2005di}, the
temperature dependence of the Casimir effect can be qualitatively different
for different geometries, as both the pure Casimir effect as well as its
thermal corrections arise from the underlying spectral properties of the
fluctuations. First analytical as well as numerical evidence of this
``geothermal'' interplay in a perpendicular-plates configuration has been
found in \cite{Gies:2008zz} using the worldline formalism.

The physical reason for this interplay can be understood in simple terms: for
the classical parallel-plate case, the nontrivial part of the fluctuation
spectrum  is given by the modes orthogonal to the plates. This relevant part
of the spectrum has a gap of wave number $k_{\text{gap}}=\pi/a$, where $a$ is
the plate separation. For small temperatures $T\ll k_{\text{gap}}$, the
higher-lying relevant modes can hardly be excited, such that their thermal
contribution to the Casimir force remains suppressed: the resulting force law
for the parallel-plates case scales like $(aT)^4$. This argument for a 
suppression of thermal contributions applies to all geometries with a gap in
the relevant part of the spectrum (e.g. concentric cylinders or spheres,
Casimir pistons, etc.). These geometries are called {\em closed}.\footnote{Of
  course, parallel plates as well as concentric cylinders are not closed in
  the sense of compactness. Also, they have a gapless part of the spectrum
  along the symmetry axes. However, this part of the spectrum does not give
  rise to the Casimir force and hence is not a relevant part.}

This reason for a suppression of thermal contributions is clearly absent for
{\em open} geometries with a relevant gapless part of the spectrum. For these
geometries, relevant modes of the spectrum can always be excited at any small
temperature value. Therefore, a stronger thermal contribution $\sim
(aT)^\alpha$ with $0<\alpha<4$ can be expected. As experimentally important
configurations such as the sphere-plate or the cylinder-plate geometry belong
to this class of open geometries, a potentially significant geothermal
interplay may exist in the relevant parameter range $aT \sim 0.01\dots0.1$. 

\begin{figure}[t]
\begin{center}
\includegraphics[width=0.95\linewidth]{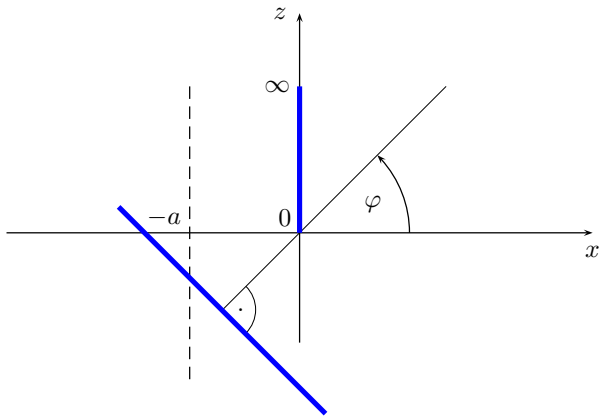}
\end{center}
\caption{Sketch of the inclined-plates configuration. The infinite plate
  (dashed line) is rotated in the $x,z$ plane by an angle $\varphi$. As
  special cases, $\varphi=0$ corresponds to the configuration of one
  semi-infinite plate parallel to an infinite plate (1si configuration),
  whereas $\varphi=\pi/2$ yields the perpendicular-plates
  configuration.}\label{T0-ip-1}
\end{figure}

In the present work, we provide further evidence for the geometry-temperature
interplay in the Casimir effect. For simplicity, we study the Casimir effect
induced by a fluctuating real scalar field obeying Dirichlet boundary
conditions ('Dirichlet scalar'). As an illustrative example, we concentrate on
an inclined-plates configuration; here, a semi-infinite plate is located above
an infinite one, with an angle of inclination of $0<\varphi\leq\pi/2$, see
Fig.~\ref{T0-ip-1}. This configuration generalizes geometries which have first
been proposed and studied in the context of Casimir edge effects
\cite{Gies:2006xe}. Our results do not only generalize the findings of
\cite{Gies:2008zz} which hold for $\varphi=\pi/2$. Most importantly, we
identify regimes with fractional temperature dependences for certain
geometries. Moreover, we work in $D=d+1$ dimensional spacetime, yielding many
analytical as well as numerical results {for the Casimir force and energy
  as well as for the torque}.

A reliable study of geothermal Casimir phenomena requires a method that is
capable of dealing with very general Casimir geometries. For this, we use the
worldline approach to the Casimir effect \cite{Gies:2003cv}, which is based on
a mapping of field-theoretic fluctuation averages onto quantum-mechanical path
integrals \cite{Feynman,Halpern:1977ru,Schmidt:1993rk}. For arbitrary backgrounds,
this worldline integral representing the spacetime trajectories of the quantum
fluctuations can straightforwardly be computed by Monte Carlo methods
\cite{Gies:2001zp}. As the computational algorithm is generally independent of
the background, i.e., the Casimir geometry in our case, Casimir problems can
straightforwardly be tackled with this method. High-precision computations for
Dirichlet-scalar fluctuations have been performed, e.g., for the sphere-plate
and cylinder-plate case \cite{Gies:2005ym,Gies:2006bt,Gies:2006cq}. In the
present work, we demonstrate that the worldline approach can also be used to
obtain novel analytical results (see also \cite{Schaden:2009zz}  for an
analytical worldline approximation technique).

In order to overcome standard approximative tools based, e.g., on the
proximity-force theorem \cite{pft1}, a variety of new field-theoretical
methods for Casimir phenomena have been developed in recent years, ranging
from improved approximation methods \cite{semicl,Scardicchio:2004fy,Balian:1977qr} to
exact methods mainly based on scattering theory
\cite{Bulgac:2005ku,Emig:2006uh,Bordag:2006vc,%
  Kenneth:2006vr,Emig:2007cf,Rodrigues:2006ku,Mazzitelli:2006ne,%
  Milton:2007gy,Milton:2008vr} or a functional integral approach
\cite{Bordag:1983zk,Emig:2001dx,Emig:2003eq}. It will certainly be worthwhile to
generalize these methods to finite temperature for a study of the
geometry-temperature interplay.

In the remainder of this introduction, we summarize our most important results
specializing to $3+1$ dimensional spacetime. In
Sect.~\ref{sec:worldl-appr-casim}, we briefly review the worldline approach to
the Casimir effect. Sect.~\ref{sec:case-t=0} is devoted to a study of the
zero-temperature Casimir effect for the geometries under consideration. The
finite-temperature case is described in Sect.~\ref{sec:finite-temperature}. Our
conclusions are summarized  in Sect.~\ref{sec:conclusions}.  

\subsection{Summary of results in $D=4$}

{Let us already summarize our most important results, specializing to
  $D=4$ spacetime dimensions and concentrating on the Casimir interaction
  energy; the corresponding force can straightforwardly be derived by
  differentiation.} At zero temperature, the classical Casimir energy of
two parallel {Dirichlet} plates at a distance $a$ reads
\begin{align}\label{int-sum-1}
\frac{E_\mathrm{c}^\parallel}{A}=-\frac{\const_\parallel \hbar c}{ a^{3}},
\ \ \ \const_\parallel=\frac{\pi^2}{1440}\approx 0.00685,  
\end{align}
where $A$ is the area of the plates. {From now on, we use natural units,
  setting} $\hbar c=1$. The Casimir energy of inclined plates {(i.p.) can be
  parameterized as}
\begin{align}\label{int-sum-2}
  \frac{E_\mathrm{c}^\mathrm{i.p.,\varphi}}{L_{y}}=&-\frac{
    \const_\varphi}{\sin(\varphi )\ a^{2}}, 
\end{align}
where the coefficient $\const_\varphi$ is shown in Fig. \ref{T0-ip-7} as a
function of $\varphi$. The extent of the inclined plate in $y$ direction
{along the edge} is $L_{y}$. At $\varphi=0$, {the energy per edge
  length} (\ref{int-sum-2}) diverges and has to be replaced by
\begin{align}\label{int-sum-3}
E_\mathrm{c}^\mathrm{1si}=E_\mathrm{c}^\mathrm{1si,\parallel}+E_\mathrm{c}^\mathrm{1si,edge}=-
\frac{A^{\mathrm{1si}} \const_\parallel }{a^3}-\frac{L_y\const_{\mathrm{edge}}}{a^2},
\end{align}
where $E_\mathrm{c}^\mathrm{1si,\parallel}$ is the Casimir Energy
(\ref{int-sum-1}) with $A^{\mathrm{1si}}$ being the semi-infinite plate's area
and $E_\mathrm{c}^\mathrm{1si,edge}$ the so-called edge {energy}. The
numerical value of $\const_{\mathrm{edge}}$ is about $0.0026$ {in
  agreement with \cite{Gies:2006xe}}.

The Casimir torque is obtained from Eq.~(\ref{int-sum-2}) by
$D_\mathrm{c}^\mathrm{i.p.,\varphi}=
\mathrm{d}E_\mathrm{c}^\mathrm{i.p.,\varphi}/\mathrm{d}\varphi$. For $D=4$,
the torque $D_\mathrm{c}^\mathrm{i.p.,\varphi}$ as a function of $\varphi$ is
shown in Fig.~\ref{CTorque-1}.  At $\varphi=0$, the Casimir torque per unit
length diverges as well but can be converted into finite torque per unit
area. Remarkably, for $\varphi=0$ the {standard} torque obtained from the
Casimir energy of parallel plates (\ref{int-sum-1})  ${D_\text{c}^\|}=A L_{z}\pi^2/960
a^4\approx0.0103 A L_{z}/ a^4$ is reduced by a {repulsive} contribution
$\approx-0.003660
L_{y} /a^2$ {arising} from the edge effect.  We encounter a similar subleading
repulsive torque effect at finite temperature.

Thermal fluctuations modify the Casimir energy, {yielding} the free energy
\begin{align}\label{int-sum-4}
E_\mathrm{c}(T)=E_\mathrm{c}(0)+\Delta E_\mathrm{c}(T),
\end{align}
where $\Delta E_\mathrm{c}(T)$ is the temperature correction.
For $(aT)\rightarrow 0$, the correction $\Delta
E^{\parallel}_\mathrm{c}(aT\rightarrow)$ to the {well-known} parallel-plates energy reads
\begin{align}\label{int-sum-5}
\frac{\Delta E^{\parallel}_\mathrm{c}(aT\rightarrow 0)}{A}=-\frac{ \zeta(3)T^3}{4 \pi}+\frac{ \pi^2 a T^4 }{90},
\end{align}
which is $\approx-0.0957 T^3+0.110 a T^4$. Note that only the $T^4$ term
contributes to the force {as the first term vanishes upon differentiation.}


For $(aT)\rightarrow 0$, {our result for} the thermal correction $\Delta
  E_\mathrm{c}^{\mathrm{i.p.},\varphi}(T)$ to the inclined-plates energy
  reads
\begin{align}\label{int-sum-6}
  \frac{\Delta E_\mathrm{c}^{\mathrm{i.p.},\varphi}(aT\rightarrow 0)}{L_{y}}
  =-\frac{ \const_{\varphi,T_0} T^2}{24 \sin(\varphi)}+\frac{ \zeta(3) a T^3}{4 \pi \sin(\varphi)}, \end{align}
where $\const_{\varphi,T_0}$ is shown in Fig. \ref{ym-integrals} as a function of
$\varphi$. The second term {which is a purely analytical result} is
{the generalization of a result for perpendicular plates,
  $\varphi=\pi/2$, found in \cite{Gies:2008zz}; numerically, this term
  evaluates to } $\approx 0.0957 a T^3/\sin(\varphi)$.

{Again, Eq.~(\ref{int-sum-6}) denotes an energy per edge length} and
diverges as $\varphi\rightarrow 0$. It has to be replaced by the formula for
the energy of a semi-infinite plate above a parallel one,
$E_\mathrm{c}^\mathrm{1si}(T)
=E_\mathrm{c}^\mathrm{1si,edge}(T)+E_\mathrm{c}^\mathrm{1si,\parallel}(T)$. The
thermal part of $E_\mathrm{c}^\mathrm{1si,\parallel}(T)$ is as in
(\ref{int-sum-5}), where $A$ is the area of the semi-infinite plate. The
leading thermal correction to the edge effect $\Delta
E_\mathrm{c}^\mathrm{1si,edge}(T)$ reads
\begin{align}\label{int-sum-7}
\frac{\Delta E_\mathrm{c}^\mathrm{1si,edge}(T)}{L_y}=-\frac{ \const_{\varphi,T_0} T^2}{24}+ 0.063 a^{1.74}  T^{3.74}, 
\end{align}
%
%
For $(aT)\rightarrow \infty$, {all thermal Casimir energies increase
  linearly in $T$ due to dimensional reduction. For instance,} the Casimir
energy $E^{\parallel}_\mathrm{c}(T)$ for parallel plates becomes
\begin{align}\label{int-sum-8}
E^{\parallel}_\mathrm{c}(aT \rightarrow \infty)=-\frac{A \zeta(3)T}{8 \pi a^2}, 
\end{align}
which is $\approx -0.0478 A T/a^2$. Note that $E^{\parallel}_\mathrm{c}(aT
\rightarrow \infty)$ is independent of $\hbar c$ as the dimensional analysis
easily shows. The energy at large $(aT)$ can therefore be interpreted as a
classical effect.

The same holds for the large $(aT)$ behavior {of the inclined-plates case as
well as for semi-infinite plates}. For inclined plates, we get
\begin{align}\label{int-sum-9}
E_\mathrm{c}^{\mathrm{i.p.},\varphi}(aT & \rightarrow 
\infty)=-\frac{L_{y}\sqrt{\pi} \const_{\varphi,T_\infty}T}{(4\pi)^{2}
a\ \sin(\varphi)}, 
\end{align}
where $\const_{\varphi,T_\infty}$ is shown in Fig. \ref{ym-integrals}   as a function of $\varphi$.

The edge effect reads  at large $(aT)$
\begin{align}\label{int-sum-10} 
E_\mathrm{c}^\mathrm{1si,edge}(aT\rightarrow \infty)=-\frac{ 0.016 L_{y} T}{a}.
\end{align}
In main part of this article, these results will be derived in detail in
$D$ spacetime dimensions.

\section{Worldline approach to the Casimir effect}
\label{sec:worldl-appr-casim}

Let us briefly review the worldline approach to the Casimir effect for a
massless Dirichlet scalar; for details, see \cite{Gies:2003cv,Gies:2006cq}.
Consider a configuration $\Sigma$ consisting of two rigid objects with
surfaces $\Sigma_1$ and $\Sigma_2$. The worldline representation of the
Casimir interaction energy in $D=d+1$ dimensional spacetime reads
\begin{align}\label{Int-1}
E_\mathrm{c}=-\frac{1}{2 (4\pi)^{D/2}} \!\int_{0}^\infty
\!\!\! \frac{\mathrm{d}\pt}{\pt^{1+D/2}} \!\int \!\mathrm{d}^d
x_\mathrm{CM}\left\langle 
\Theta_\Sigma [\mathbf{x}(\tau)] \right\rangle.
\end{align}
Here, the generalized step functional obeys $\Theta_\Sigma[\mathbf{x}]=1$
if a worldline $\mathbf{x}(\tau)$ intersects both surfaces
$\Sigma=\Sigma_1\cup\Sigma_2$, and is zero otherwise. 

The expectation value in \Eqref{Int-1} is taken with respect to an ensemble of
$d$-dimensional closed worldlines with a common center of mass
$\mathbf{x}_{\mathrm{CM}}$ and a Gau\ss ian velocity distribution,
\begin{equation}
\langle \dots \rangle = \frac
{  \int_{\mathbf{x}_{\text{CM}}} \mathcal{D} \mathbf x\,\dots\,  e^{-\frac{1}{4}
  \int_0^{\pt} \mathrm{d}\tau \, \dot{\mathbf x}^2(\tau)}}
{
  \int_{\mathbf{x}_{\text{CM}}} \mathcal{D} \mathbf x \, e^{-\frac{1}{4}
  \int_0^{\pt} \mathrm{d}\tau \, \dot{\mathbf x}^2(\tau)}}.\label{eq:expval}
\end{equation}
Here, we have already used the fact that the time component cancels out for
static Casimir configurations at zero temperature.  Eq.~(\ref{Int-1}) has an
intuitive interpretation: All worldlines intersecting both surfaces do not
satisfy Dirichlet boundary conditions on both surfaces. They are removed from
the ensemble of allowed fluctuations by the $\Theta$ functional and thus
contribute to the negative Casimir interaction energy. In the process of the
auxiliary $\pt$ integration, the propertime parameter $\pt$
scales the extent of a worldline by a factor of $\sqrt{\pt}$. Large
$\pt$ correspond to long-wavelength or IR fluctuations, small
$\pt$ to short-wavelength or UV fluctuations.

Introducing finite temperature $T=1/\beta$ by the Matsubara formalism {is
  equivalent} to compactifying {Euclidean} time on the interval
$[0,\beta]$. Now, the closed worldlines live on a cylindrical surface and can
carry a winding number. The worldlines $\mathbf{x}^{(n)}(\tau)$ winding $n$
times around the cylinder can be decomposed into a worldline $\tilde{\mathbf
  x}(\tau)$ with no winding number and a winding motion at constant speed,
\begin{align}\label{Int-2}
x_i^{(n)}(\tau)=\tilde{x}_i
(\tau)+\frac{n\beta\tau}{\pt}\delta_{iD},
\end{align}
where the $D$th component corresponds to Euclidean time. The Casimir energy
(\ref{Int-1}) now becomes
\begin{eqnarray}\label{Int-3}
E_\mathrm{c}&=&-\frac{1}{2 (4\pi)^{D/2}} 
\\
&&\times \! \! \int_{0}^\infty\!\!\!
\frac{\mathrm{d}\pt}{\pt^{1+D/2}}\!\!\sum_{n=-\infty}^{\infty} \!\!e^{-\frac{n^2\beta^2}{4
\pt}}\!\! \int \! \mathrm{d}^d x_\mathrm{CM}\left\langle
\Theta_\Sigma [\mathbf{x}(\tau)] \right\rangle.\nonumber
\end{eqnarray}
The finite-temperature worldline formalism for static configurations thus
boils down to a winding-number prefactor in front of the worldline expectation
value together with a sum over winding numbers:
\begin{align}\label{Int-4}
\langle\dots\rangle \ \rightarrow \ \left(1+2\sum_{n=1}^\infty
e^{-\frac{n^2\beta^2}{4\pt}}\right)\langle\dots\rangle.
\end{align}
The winding-number sum is directly related to the standard Matsubara sum by a
Poisson resummation,
\begin{align}\label{Int-4b}
  \left(1+2\sum_{n=1}^\infty e^{-\frac{n^2\beta^2}{4\pt}}\right) 
  = \frac{\sqrt{4 \pi \pt}}{\beta} \sum_{n=-\infty}^\infty e^{- \left(
    \frac{2\pi m}{\beta} \right)^2 \pt}. 
\end{align}
This is already sufficient to understand the high-temperature limit of generic
Casimir configurations: at high temperatures $\beta\to0$, only the zeroth
Matsubara frequency survives as higher modes receive thermal masses of order
$\sim 2\pi/\beta=2\pi T$ and decouple. All remaining temperature dependence
arises from the dimensional prefactor $1/\beta ={T}$, and the dependence on
the Casimir geometry only enters the prefactor. The calculation of the latter
is a dimensionally reduced problem in $D-1$ dimensions. This is a general
mechanism of {\em dimensional reduction} in high-temperature field theories.
The linear high-temperature asymptotics is also clear from the fact that the
Bose-Einstein distribution governing the distribution of bosonic thermal
fluctuations increases as $\sim {T}$ in the high-temperature limit.

Finally, it is advantageous for numerical as well as analytical calculations
to rescale the worldlines such that the velocity distribution becomes
independent of $\pt$,
\begin{equation}
\bgam(t):= \frac{1}{\sqrt{\pt}} \mathbf x(\pt t)\quad \!\!\!\to\!\!\!\quad
e^{-\frac{1}{4} \int_0^{\pt}   \dot{\mathbf{x}}^2 \mathrm{d}\tau} = e^{-\frac{1}{4}
  \int_0^1  \dot{\bgam}^2 \mathrm{d}t},
\end{equation}
where the dot always denotes a derivative with respect to the argument, e.g.,
$\dot{\bgam}=\mathrm{d}\bgam(t)/\mathrm{d}t$. In terms of these
normalized worldlines $\bgam$ and the center-of-mass coordinate $\mathbf{x}_{\text{CM}}$,
the $\Theta$ function reads more explicitly
\begin{equation}
\Theta[\mathbf x]\equiv\Theta[\mathbf
x_{\text{CM}}+\sqrt{\pt}\bgam(t)].
\end{equation}
The involved worldline integrals can be evaluated also numerically by Monte
Carlo methods in a straightforward manner. For this, the path integral over an
operator $\mathcal O$ is approximated by a sum over a finite ensemble of $\nL$
loops,
\begin{equation}
  \langle \mathcal O[\bgam] \rangle\, \to\, \frac{1}{\nL} \sum_{\ell=1}^{\nL} \mathcal O[\bgam_\ell],
\end{equation}
where $\ell$ counts the worldlines in the ensemble. Each worldline $\bgam(t)$
is furthermore discretized by a finite set of $N$ points per loop (ppl),
\begin{equation}\label{intro-10}
  \bgam(t)\, \to  \,\bgam_i=\bgam(t_i), \quad t_i= \frac{i}{N}, \,\,\,
  i=0,\dots, N,
\end{equation}
where $\bgam_0=\bgam_N$ are identified as the worldlines are closed. Various
efficient ab initio algorithms for generating discretized worldlines with
Gau\ss ian velocity distribution have been developed, see, e.g.,
\cite{Gies:2003cv,Gies:2005sb}.

With these comparatively simple prerequisites, we can now
turn to an analysis of various non-trivial Casimir configurations for the
Dirichlet scalar.

\section{Casimir effect at zero temperature}
\label{sec:case-t=0}

Let us first study parallel and inclined plates at zero temperature. The
purpose of this section is on the one hand to review and generalize known
results and on the other hand to exemplify how the Casimir effect can be
understood in terms of simple geometric properties of the worldlines. 

\subsection{Parallel Plates}
We start with Casimir's classic configuration of two infinitely extended
parallel plates. Let the lower and upper plate lie in the $z=-a$ and $z=0$
planes, respectively. In $d$ space dimensions the surface area $A$ of the
plates is then $d-1$ dimensional. The $\Theta$ functional for this
configuration reads
\begin{figure}[t]
\begin{center}
\includegraphics[width=0.95\linewidth]{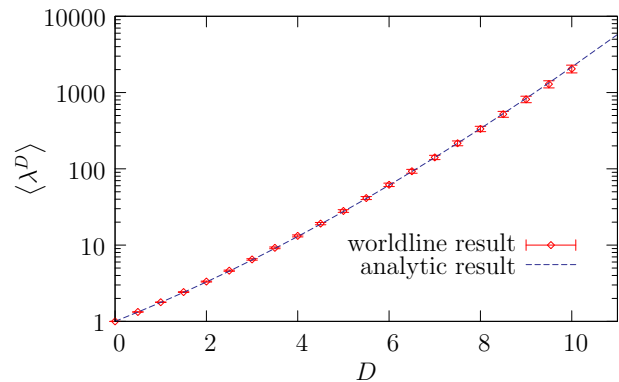}
\end{center}
\caption{$D$th moment of the maximum spatial extent $\lambda$ of a worldline
  as a function of $D$. This geometric object (which is the same in any target
  dimension of the worldline) is related to the Casimir energy of the
  parallel-plates configuration in $D$ spacetime dimensions by
  Eqs. \eqref{T0-pp-3} and \ref{T0-pp-4}. The plot compares the exact
  analytical result with the worldline numerical computation based on 1000
  worldlines with $2\times10^6$ ppl (points per loop) each.}\label{T0-pp-5}
\end{figure}
\begin{align}\label{T0-pp-1}
\Theta_\parallel\left[z_\mathrm{CM}+\sqrt{\pt} \gamma_{z,\ell}\right]=&\,\theta\left(\sqrt{\pt}
\gamma_{{z_{\mathrm{max}}},\ell}+z_\mathrm{CM}\right)
\\
&\times\theta\left(-z_\mathrm{CM}-\sqrt{\pt}
\gamma_{{z_{\mathrm{min}}},\ell}-a\right),\notag
\end{align}
where $\gamma_{z,\ell}$ is the $z$ coordinate of the $\ell$'th worldline
(measured with respect to the center of mass). The quantities
$\gamma_{{z_{\mathrm{max}}},\ell}, \ \gamma_{{z_{\mathrm{min}}},\ell}$ denote
the worldline's maximal and minimal extent in the $z$ direction,
respectively. The total maximal extent $\lambda_\ell$ of the $\ell$'th
worldline then is $\lambda_\ell=
\gamma_{{z_{\mathrm{max}}},\ell}-\gamma_{{z_{\mathrm{min}}},\ell}$. Now, we
can do the integral in \Eqref{Int-1} and obtain the Casimir energy density
(suppressing the index $\ell$ from now on)
\begin{align}\label{T0-pp-2}
\frac{E_\mathrm{c}^\parallel}{A}=-\frac{\langle \lambda^D\rangle}{ D(D-1)(4\pi)^{D/2}\ a^{D-1}}.
\end{align}
We observe that the $D$-dimensional parallel-plate Casimir energy is related
to the $D$th cumulant of the extent of the worldlines \cite{Gies:2006cq}. This
is a first example for a relation between Casimir energies and geometric
properties of the worldlines. Instead of computing these cumulants directly,
let us simply compare \Eqref{T0-pp-2} with the well-known analytic result
\cite{Verschelde:1985jr,Svaiter:1989gz}.
\begin{align}\label{T0-pp-3}
\frac{E_\mathrm{c}^\parallel}{A}=-\frac{\Gamma(D/2)\zeta (D)}{
(4\pi)^{D/2}\ a^{D-1}},
\end{align}
yielding
\begin{align}\label{T0-pp-4}
\langle \lambda^D\rangle=D(D-1)\Gamma(D/2)\zeta (D).
\end{align}
A comparison of the analytical result to a numerical evaluation of the
cumulants is displayed in Fig. \ref{T0-pp-5}.  Also, the Casimir force density
can straightforwardly be obtained as the derivative of \Eqref{T0-pp-2} with
respect to $a$.  Incidentally,  the connection between Casimir energies and
worldline properties also induces a relation between Casimir energies and
questions in polymer physics, as first observed in \cite{Gies:2006cq}.

\subsection{Inclined plates}

\begin{figure}[t]
\begin{center}
\includegraphics[width=0.95\linewidth]{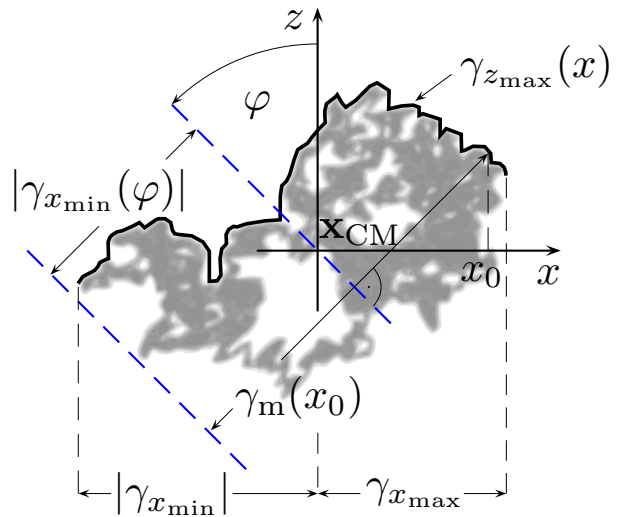}
\end{center}
\caption{All relevant information for the evaluation of the Casimir energy of
  the inclined plates, (\ref{T0-ip-6}), is encoded in the function
  $\gamma_\mathrm{m}(x)$, which has to be integrated from
  $\gamma_{x_\mathrm{min}}$ to $\gamma_{x_\mathrm{max}}$.}\label{T0-ip-iloop}
\end{figure}

The inclined-plates (i.p.) configuration consists of a perfectly thin semi-infinite
plate above an infinite plate at an angle $\varphi$, see
Fig. \ref{T0-ip-1}. The semi-infinite plate has an edge with a ($d-2$
dimensional) length $L_{y}$. The infinite plate has a ($d-1$) dimensional area
$A$.\footnote{Of course, the labels ``semi-infinite'' and ``infinite'' imply
  that both $L_y$ and $A$ are considered in the limit $L_y,A\to\infty$.} Let
$a$ be the minimal distance between the plates. In the following, we will omit
the center-of-mass subscript CM. The $\Theta$ functional for this
configuration reads
\begin{align}\label{T0-ip-2}
  \Theta_{\mathrm{i.p.},\varphi}=&
  \theta\left(-x\cos(\varphi)-z\sin(\varphi)
  -\sqrt{\pt}\gamma_{{x_{\mathrm{min}}},\ell}(\varphi)-a\right) \notag
  \\
  &\times\theta\left(z+\sqrt{\pt}\gamma_{{z_{\mathrm{max}}},\ell}
  \left(-\frac{x}{\sqrt{\pt}}\right)\right)
  \\
  &\times\theta\left(-x-\sqrt{\pt}\gamma_{{x_{\mathrm{min}}},\ell}\right)
  \theta\left(x+\sqrt{\pt}\gamma_{{x_{\mathrm{max}}},\ell}\right),\notag 
\end{align}
where the first $\theta$ function ensures the intersection of the worldline
with the infinite plate. The remaining three ones account for the intersection
with the semi-infinite plate. In \Eqref{T0-ip-2}, we have used
\begin{align}\label{T0-ip-3}
  \gamma_{{x_{\mathrm{min}}},\ell}(\varphi)&
  \equiv\min_t\left(\gamma_{{x},\ell}(t)\cos(\varphi)
  +\gamma_{z,\ell}(t)\sin(\varphi)\right), 
\end{align}
where $t$ parameterizes the worldline; i.e., in the discretized version, we
have $t=1\dots N$ with $N$ being the number of points per worldline loop (ppl).
Trivially,
$\gamma_{{x_{\mathrm{min}}},\ell}(0)=\gamma_{{x_{\mathrm{min}}},\ell}$ and
$\gamma_{{x_{\mathrm{min}}},\ell}(\pi/2)=\gamma_{{z_{\mathrm{min}}},\ell}$
holds. In other words, $\gamma_{{x_{\mathrm{min}}},\ell}(\varphi)$ measures
the minimal extent of the worldline in the $x$ direction of a coordinate
system rotated by the angle $\varphi$. {In Eq.~\eqref{T0-ip-2}, we also
  encounter $\gamma_{z_{\text{max}},\ell}(x)$, denoting the $x$-dependent
  envelope of the worldline in positive $z$ direction. All these geometric
  properties of a worldline are displayed in Fig.~\ref{T0-ip-iloop}.}

The $\Theta_{\mathrm{i.p.},\varphi}$ functional in \Eqref{T0-ip-2} generalizes
the case of perpendicular plates ($\bot$) for $\varphi=\pi/2$ and the case of
one semi-infinite plate parallel to a infinite one (1si) for $\varphi\to0$; both edge
configurations were studied in detail in \cite{Gies:2006xe,Klingmuller:2007}.

Let us define
\begin{align}\label{T0-ip-5b}
\gamma_\mathrm{m}(x)\equiv x \cos (\varphi )+\sin
(\varphi ) \gamma_{z_{\mathrm{max}}}(x)-\gamma_{x_{\mathrm{min}}}(\varphi).
\end{align}
Inserting $\Theta_{\mathrm{i.p.},\varphi}$ for $\varphi\neq 0$ into
\Eqref{Int-1} leads to the Casimir energy density of the inclined plates
\begin{figure}[t]
\begin{center}
\includegraphics[width=0.95\linewidth]{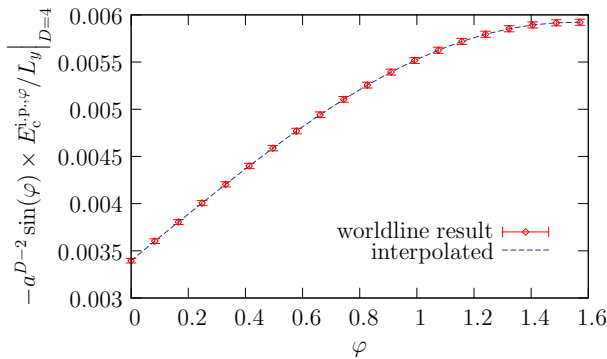}
\end{center}
\caption{Normalized Casimir energy per edge length $-\frac{
    E_\mathrm{c}^\mathrm{i.p.,\varphi}}{L_{y}}\times a^{D-2}\sin(\varphi)$ of
  the inclined-plates (i.p.) configuration in $D=4$ versus the angle of
  inclination $\varphi$. For $\varphi= 0$, this function can be evaluated
  analytically, yielding $\pi^2/2880\approx 3.427\cdot 10^{-3}$. We have used
  $20000$ worldlines with $10^6$ ppl each.} \label{T0-ip-7}
\end{figure}
\begin{align}
  \frac{E_\mathrm{c}^\mathrm{i.p.,\varphi}}{L_{y}}=&-\frac{\csc (\varphi )}{
    (4\pi)^{D/2}(D-1)(D-2) \ a^{D-2}} \notag
  \\
  & \times\left\langle
  \int_{\gamma_{x_\mathrm{min}}}^{\gamma_{x_\mathrm{max}}} \mathrm{d} x \ 
\gamma_\mathrm{m}^{D-1}(x)\right\rangle.
\label{T0-ip-6}
\end{align}
Equation (\ref{T0-ip-6}) is shown as a function of $\varphi$ in
Fig. \ref{T0-ip-7} for $D=4$.  For $\varphi= \pi/2$ and $D=4$, we rediscover
the perpendicular plates result \cite{Gies:2006xe,Klingmuller:2007} as a special case. Incidentally,
the integral in \Eqref{T0-ip-6} can be done analytically for $\varphi=0$
resulting in $\langle\lambda^D/D\rangle=(D-1)\Gamma(D/2)\zeta(D)$.  Together
with the $\varphi$-dependent prefactor, \Eqref{T0-ip-6} diverges as
$\varphi\rightarrow 0$ as it should. This is because \Eqref{T0-ip-6}
corresponds to the energy per unit \textit{edge length}, whereas for
$\varphi\to0$ the Casimir energy becomes proportional to the {\em area} of the
semi-infinite plate.  We devote the whole next section to analyzing how the
limit $\varphi\to 0$ yielding the 1si configuration can be obtained.

\subsection{Inclined plates, $\varphi\rightarrow 0$ limit}
\label{sec:incl-plat-varph}

It is instructive to study the limit of a semi-infinite plate parallel to an
infinite plate (1si), $\varphi\to 0$, as it involves a subtle limiting
process.  Recalling the general considerations of \cite{Gies:2006xe, Klingmuller:2007}
for the 1si case, the total Casimir interaction energy decomposes into
\begin{align}\label{T0-ip-4}
  E_\mathrm{c}^\mathrm{1si}
  =E_\mathrm{c}^\mathrm{1si,\parallel}+E_\mathrm{c}^\mathrm{1si,edge},
\end{align}
where $E_\mathrm{c}^\mathrm{1si,\parallel}/A$ is the usual Casimir energy per
unit area of two parallel plates \Eqref{T0-pp-3}, with $A$ being now the area
of the semi infinite plate. The so called edge energy
$E_\mathrm{c}^\mathrm{1si,edge}$ measures the contribution that arises solely
due to the presence of the edge. 

In the limit $\varphi\to0$, this decomposition is naturally achieved by
inserting $\Theta_{\mathrm{i.p.},\varphi=0}$ of \Eqref{T0-ip-2} into
\Eqref{Int-1} and performing the $z$ integral first. This leads to
\begin{align}\label{T0-ip-5}
  \frac{E_\mathrm{c}^\mathrm{1si,edge}}{L_{y}}
  =&-\frac{1}{(4\pi)^{D/2}(D-2)a^{D-2}}
  \\
  &\times \left\langle
  \int_{\gamma_{x_{\mathrm{min}}}}^{\gamma_{x_{\mathrm{max}}}} \mathrm{d}
  x\ \gamma_{z_{\mathrm{max}}}(x) 
  (x-\gamma_{x_{\mathrm{min}}})^{D-2}\right\rangle.\notag
\end{align}
This representation can straightforwardly be computed numerically
\cite{Gies:2006xe}. Of course, for truly infinite plates, the edge effect
being proportional to the length of the edge is completely negligible in
comparison with $E_\mathrm{c}^\mathrm{1si,\parallel}$, the latter being
proportional to the area of the plates. However, dealing with finite plates,
the edge effect contributes to the Casimir force, effectively increasing the
plate's area \cite{Gies:2006xe}.

Of course, the same result has to arise from the general inclined-plates
formula \Eqref{T0-ip-6} in the limit $\varphi\to0$. However, this
representation naively exhibits a divergence in this limit. To find the origin
of the divergence, we decompose \Eqref{T0-ip-6} into the parts corresponding
to the edge effect $E_\mathrm{c}^\mathrm{edge,\varphi}$ and the
semi-infinite-plates energy $E_\mathrm{c}^\mathrm{\parallel,\varphi}$,
characterized by the integrals $\int_{-\gamma_\mathrm{zmax(x)}}^0 \dots dz$
and $\int_0^{L_z/2} \dots dz$, respectively .  Here $L_z/2$ denotes the
(infinite) length of the semi-infinite plate in $z$ direction.  The result for
$E_\mathrm{c}^\mathrm{edge,\varphi}$ reads
\begin{align}\label{T0-ip-l-1}
  E_\mathrm{c}^\mathrm{edge,\varphi}=&-\frac{  L_{y} }{(4\pi)^{D/2} (D-2)a^{D-2}}\times \notag
  \\
  &
  \times\left\langle\int_{\gamma_{x_\mathrm{min}}}^{\gamma_{x_\mathrm{max}}} 
  \gamma_{z_{\mathrm{max}}}(x) \gamma_\mathrm{m}^{D-2}(x)\mathrm{d}x\right\rangle,
\end{align}
which becomes $E_\mathrm{c}^\mathrm{edge}$ in \Eqref{T0-ip-5} as
$\varphi\rightarrow 0$; Eq. (\ref{T0-ip-l-1}) is therefore valid for
$\varphi=0$.

On the other hand, if we naively expand the result for
$E_\mathrm{c}^\mathrm{\parallel,\varphi}$ for small $\varphi$, we obtain
\begin{align}\label{T0-ip-l-2}
  E_\mathrm{c}^\mathrm{\parallel}\stackrel{?}{=}\frac{ -\langle
    \lambda^D\rangle L_{y} }{(4\pi) ^{D/2}(D-2) (D-1) D a^{D-2}\varphi
  }+O(\varphi),
\end{align}
which is only valid for $\varphi\neq 0$ and does not reproduce \Eqref{T0-pp-2}
in the limit $\varphi\to0$. Instead of the energy per area, we have obtained
the energy per length, which of course diverges in this limit. In order to
rediscover the Casimir energy for the 1si configuration, the limits
$\varphi\to 0$ and the implicit limit $L_z\to\infty$ have to be taken in the
right order. In \Eqref{T0-ip-l-2}, the limit $L_z\to\infty$ has implicitly
been performed first, which precisely leads to the divergence of the energy
per edge length. Therefore, we need to first perform the limit $\varphi\to0$
at finite $L_z$ in order to obtain the desired energy per area. Starting from
$E_\mathrm{c}^\mathrm{\parallel,\varphi}$ at small $\varphi$,
\begin{align}\label{T0-ip-l-3}
  &E_\mathrm{c}^\mathrm{\parallel,\varphi\rightarrow 0}
  =-\frac{L_{y}}{2 (4 \pi)^{D/2}}\left\langle\int_0^{\infty} 
  \frac{\mathrm{d}\mathcal{T}}{\mathcal{T}^{(D+1)/2}}\int_0^{L_z/2}
  \mathrm{d}z \right.\notag
\\
&\left. \times\int_{\gamma_{x_\mathrm{min}}}^{\gamma_{x_\mathrm{max}}}\mathrm{d}x
\theta \left(-a+\sqrt{\mathcal{T}} x -z \varphi-\sqrt{\mathcal{T}}
\gamma_{x_\mathrm{min}} \right)\right\rangle, 
\end{align}
we do the $\mathcal{T}$ integral \textit{first} and obtain
\begin{align}\label{T0-ip-l-4}
  E_\mathrm{c}^\mathrm{\parallel,\varphi\rightarrow 0}=-\frac{L_{y}\langle
    \lambda^D\rangle (a^{2-D}-( a+L_z \varphi/2 )^{2-D})}{(4 \pi)^{D/2} (D-2)
    (D-1) D \varphi}. 
\end{align}
For small $(L_z \varphi)$, i.e., finite $L_z$ and $\varphi\to 0$, we can
expand the last factor in $\varphi$,
\begin{align}\label{T0-ip-l-5}
  E_\mathrm{c}^\mathrm{\parallel,\varphi\rightarrow 0}
  \cong&-\frac{L_{y} \langle  \lambda^D\rangle}{(4 \pi)^{D/2}(D-1)D}\notag
\\
&\times\left(\frac{1}{2} a^{1-D} L_z-\frac{1}{8} a^{-D}
   (D-1) L_z^2 \varphi\right),
\end{align}
which for $\varphi\equiv 0$ corresponds exactly to the parallel-plates
contribution $E_\mathrm{c}^\mathrm{1si,\parallel}$ in \Eqref{T0-pp-2}.  From
\Eqref{T0-ip-l-4}, we also observe that the other order of limits, taking
first $L_z\to \infty$ while keeping $\varphi$ finite, reproduces the divergent
behavior of the energy per edge length in \Eqref{T0-ip-l-2} (as long as
Re$[D]>2$). The proper order of limits is similarly important at finite
temperature with the additional complication that another dimensionful
parameter occurs.

\subsection{{Casimir torque of inclined plates}}

\begin{figure}[t]
\begin{center}
\scalebox{0.7}{
\begin{picture}(0,0)%
\includegraphics{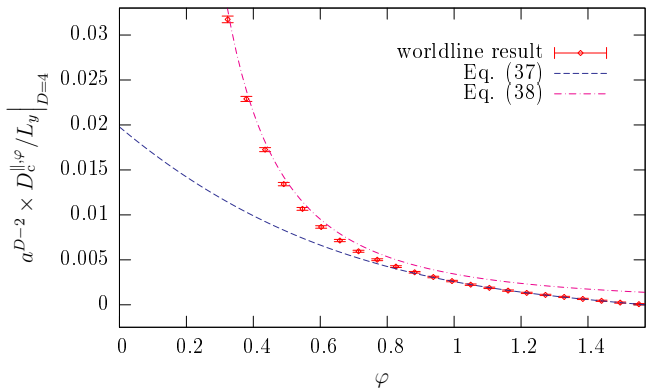}%
\end{picture}%
\begingroup
\setlength{\unitlength}{0.0200bp}%
\begin{picture}(18000,10800)(0,0)%
\large
\put(2750,2256){\makebox(0,0)[r]{\strut{} 0}}%
\put(2750,3467){\makebox(0,0)[r]{\strut{} 0.005}}%
\put(2750,4678){\makebox(0,0)[r]{\strut{} 0.01}}%
\put(2750,5889){\makebox(0,0)[r]{\strut{} 0.015}}%
\put(2750,7101){\makebox(0,0)[r]{\strut{} 0.02}}%
\put(2750,8312){\makebox(0,0)[r]{\strut{} 0.025}}%
\put(2750,9523){\makebox(0,0)[r]{\strut{} 0.03}}%
\put(3025,1100){\makebox(0,0){\strut{} 0}}%
\put(4827,1100){\makebox(0,0){\strut{} 0.2}}%
\put(6628,1100){\makebox(0,0){\strut{} 0.4}}%
\put(8430,1100){\makebox(0,0){\strut{} 0.6}}%
\put(10232,1100){\makebox(0,0){\strut{} 0.8}}%
\put(12033,1100){\makebox(0,0){\strut{} 1}}%
\put(13835,1100){\makebox(0,0){\strut{} 1.2}}%
\put(15636,1100){\makebox(0,0){\strut{} 1.4}}%
\put(550,5950){\rotatebox{90}{\makebox(0,0){\strut{}$a^{D-2}\times D_\mathrm{c}^\mathrm{\parallel,\varphi}/L_y\Big|_{D=4}$}}}%
\put(10100,275){\makebox(0,0){\strut{}$\varphi$}}%
\put(14461,9039){\makebox(0,0)[r]{\strut{}worldline result}}%
\put(14461,8489){\makebox(0,0)[r]{\strut{}Eq. (\ref{T0-ip-torque-2})}}%
\put(14461,7939){\makebox(0,0)[r]{\strut{}Eq. (\ref{T0-ip-torque-3})}}%
\normalsize
\end{picture}%
\endgroup
}
\end{center}
\caption{{Normalized Casimir torque per edge length ${\frac{D_\mathrm{c}^\mathrm{i.p.,\varphi}}{L_{y}}}\times a^{D-2}$ of
  the inclined-plates configuration in $D=4$ and its expansion around $\varphi=\pi/2$ and $\varphi=0$, respectively, versus the angle of
  inclination $\varphi$. We have used
  $10000$ worldlines with $10^6$ ppl each.}} \label{CTorque-1}
\end{figure}

The Casimir torque ${D_\mathrm{c}^\mathrm{i.p.,\varphi}}$ {referring
  to rotations of one of the plates about the edge axis} can easily be
obtained by taking the derivative of the Casimir energy (\ref{T0-ip-6}), (or
Eq.~(\ref{T0-ip-l-4}) for small $\varphi$), with respect to the angle of
inclination:
\begin{align} \label{T0-ip-torque-1}
{D_\mathrm{c}^\mathrm{i.p.,\varphi}}=
\frac{\mathrm{d}E_\mathrm{c}^\mathrm{i.p.,\varphi}}{\mathrm{d}\varphi}. 
\end{align}
For $\varphi$ near $\pi/2$, we can even set $\mathrm{d}\gamma_{x_\mathrm{min}}(\varphi)/\mathrm{d}\varphi=0$ before
  taking the average with respect to the loop ensemble, simplifying the
  calculations. This is, because the derivative
  $\mathrm{d}\gamma_{x_\mathrm{min}}(\varphi)/\mathrm{d}\varphi$ changes its sign
  for perpendicular plates $\varphi=\pi/2$ if the worldline is rotated by
    an angle $\pi$ about the normal axis of the lower plate, see
  Fig. \ref{T0-ip-1}. Therefore, the sign correlates with the position of the
  minimum on the $x$ axis of the lower plate. But the position $x$ of the
  minimum $\gamma_{x_\mathrm{min}}(\varphi)$ does not correlate with the value of
  the integral in Eq.~(\ref{T0-ip-6}) leading to a mutual cancellation of terms
  involving $\mathrm{d}\gamma_{x_\mathrm{min}}(\varphi)/\mathrm{d}\varphi$.  

For $\varphi<\pi/2$, we have to rotate the worldline about the normal axis of
the inclined lower plate. Then, the correlation between the position of the
minimum and the involved integrals does not vanish any more since the original
and rotated worldline contribute differently to the integral.  In general,
expressions containing derivatives of $\gamma_{x_\mathrm{min}}(\varphi)$ cannot
be neglected even at $\varphi=\pi/2$. Since the worldlines are not smooth, the
convergence of averages of such expressions will be very slow. This is the
case when calculating the coefficients of an expansion of Eq.~(\ref{T0-ip-torque-1})
near $\varphi=\pi/2$. Since the second derivative already appears in the first
expansion coefficient, more confident values are obtained by a numerical fit
to Eq.~(\ref{T0-ip-torque-1}). There, only the first derivative is
present. For $D=4$, we obtain (see Fig. \ref{CTorque-1})
\begin{align} \label{T0-ip-torque-2}
\frac{{D_\mathrm{c}^\mathrm{i.p.,\varphi\rightarrow \pi/2}} a^2}{L_y}\approx
0.00329\left(\frac{\pi}{2}-\varphi\right)+0.0038\left(\frac{\pi}{2}-\varphi\right)^3. 
\end{align}
This should be compared to the worldline average based on the expansion of
Eq.~(\ref{T0-ip-torque-1}) around $\pi/2$: the linear coefficient in
Eq.~(\ref{T0-ip-torque-2}) then yields 
0.003 $\pm$ 0.0002
. If we
neglect all derivatives of $\gamma_{x_\mathrm{min}}(\varphi)$ the worldline
result reads 
0.00285 $\pm$ 0.00003
. In all three cases $10000$
worldlines with $10^6$ ppl were used.

For $\varphi\rightarrow 0$, the Casimir torque diverges. The expansion about
$\varphi=0$ can easily be obtained analytically from (\ref{T0-ip-l-4})
\begin{align} \label{T0-ip-torque-3}
  {D_\mathrm{c}^\mathrm{i.p.,\varphi\rightarrow 0}} \cong
\frac{L_{y}\Gamma(D/2)\zeta (D) }{(4 \pi)^{D/2} (D-2)
     a^{D-2}\varphi^2},
\end{align}
where we have used Eq.~(\ref{T0-pp-4}). For $D=4$, Eq.~(\ref{T0-ip-torque-3})
yields $L_y\pi^2/2880 a^2\varphi^2\approx0.00343 L_y/a^2\varphi^2$, being
excellent approximation to Eq.~(\ref{T0-ip-torque-1}) for $\varphi$ not too close
to $\pi/2$.

The divergent Casimir torque per length can be converted into finite torque per
unit area by means of Eq.~(\ref{T0-ip-l-4}). Note that Eq.~(\ref{T0-ip-l-4}) leads
to the classical result for the torque,
\begin{align} \label{T0-ip-torque-4}
  D_\mathrm{c}^\mathrm{\parallel,\varphi\rightarrow 0} =
\frac{A L_{z}\Gamma(D/2)\zeta (D) (D-1) }{2(4 \pi)^{D/2}
     a^{D}},
\end{align}
where $A$ and $L_z$ denote the semi-infinite plate's area and extent in $z$
direction, respectively. For $D=4$, Eq.~(\ref{T0-ip-torque-4}) becomes $A
L_{z}\pi^2/960 a^4\approx0.0103 A L_{z}/ a^4$. 

\begin{figure}[t]
\begin{center}
\scalebox{0.7}{
\begin{picture}(0,0)%
\includegraphics{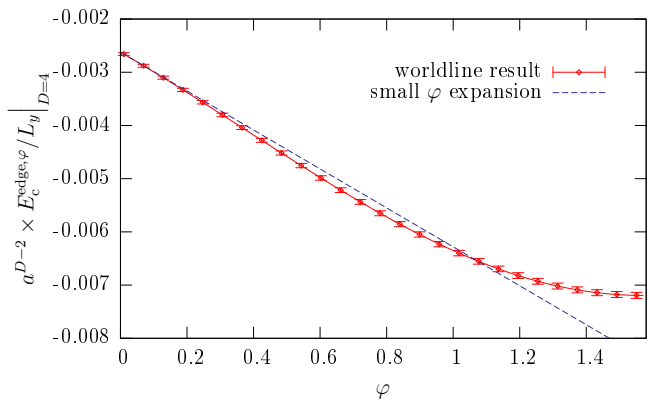}%
\end{picture}%
\begingroup
\setlength{\unitlength}{0.0200bp}%
\begin{picture}(18000,10800)(0,0)%
\large
\put(2750,1650){\makebox(0,0)[r]{\strut{}-0.008}}%
\put(2750,3083){\makebox(0,0)[r]{\strut{}-0.007}}%
\put(2750,4517){\makebox(0,0)[r]{\strut{}-0.006}}%
\put(2750,5950){\makebox(0,0)[r]{\strut{}-0.005}}%
\put(2750,7383){\makebox(0,0)[r]{\strut{}-0.004}}%
\put(2750,8817){\makebox(0,0)[r]{\strut{}-0.003}}%
\put(2750,10250){\makebox(0,0)[r]{\strut{}-0.002}}%
\put(3025,1100){\makebox(0,0){\strut{} 0}}%
\put(4816,1100){\makebox(0,0){\strut{} 0.2}}%
\put(6607,1100){\makebox(0,0){\strut{} 0.4}}%
\put(8398,1100){\makebox(0,0){\strut{} 0.6}}%
\put(10190,1100){\makebox(0,0){\strut{} 0.8}}%
\put(11981,1100){\makebox(0,0){\strut{} 1}}%
\put(13772,1100){\makebox(0,0){\strut{} 1.2}}%
\put(15563,1100){\makebox(0,0){\strut{} 1.4}}%
\put(550,5950){\rotatebox{90}{\makebox(0,0){\strut{}$ a^{D-2}\times E_\mathrm{c}^\mathrm{edge,\varphi}/L_{y}\Big|_{D=4} $}}}%
\put(10100,275){\makebox(0,0){\strut{}$\varphi$}}%
\put(14392,8817){\makebox(0,0)[r]{\strut{}worldline result}}%
\put(14392,8267){\makebox(0,0)[r]{\strut{}small $\varphi$ expansion}}%
\normalsize
\end{picture}%
\endgroup
}
\end{center}
\caption{{Normalized edge energy per edge length $\frac{E_\mathrm{c}^\mathrm{edge,\varphi}}{L_{y}}\times a^{D-2}$ of
  the inclined-plates configuration in $D=4$ and its expansion around $\varphi=0$ versus the angle of
  inclination $\varphi$. We have used
  $10000$ worldlines with $10^6$ ppl each.}} \label{T0-ip-torque-6}
\end{figure}

A new characteristic contribution emerges from the edge effect
Eq.~(\ref{T0-ip-l-1}). Unlike the total inclined-plate Casimir energy
$E_\mathrm{c}^\mathrm{i.p.,\varphi}=E_\mathrm{c}^\mathrm{\parallel,\varphi}+E_\mathrm{c}^\mathrm{edge,\varphi}$,
the edge energy (\ref{T0-ip-l-1}) decreases with the angle of inclination
$\varphi$, see Fig. \ref{T0-ip-torque-6}. This leads to a contribution which
works against the standard torque (\ref{T0-ip-torque-4}). For $D=4$, the
correction to Eq.~(\ref{T0-ip-torque-4}) emerging from the edge effect reads
\begin{align} \label{T0-ip-torque-5}
  D_\mathrm{c}^\mathrm{edge,\varphi\rightarrow 0} =
-(0.003660 \pm 0.000038)\frac{ L_{y} }{a^2},
\end{align}
%
where we have used $10000$ worldlines with $10^6$ ppl each.   The coefficient
  in Eq.~(\ref{T0-ip-torque-5}) was calculated by expanding 
{Eq.~(\ref{T0-ip-l-1})}
 around $\varphi=0$. Equation (\ref{T0-ip-torque-5}) is shown in
  Fig. \ref{T0-ip-torque-6}. We will see a similar subleading repulsive torque effect in
  the next section where we investigate finite-temperature contributions.

\section{Finite temperature}
\label{sec:finite-temperature}

Decomposing the Casimir energy at finite temperature $T=1/\beta$ into its
zero-temperature part $E_\mathrm{c}(0)$ and finite-temperature correction
$\Delta E_\mathrm{c}(T)$,
\begin{align}\label{FT-1}
E_\mathrm{c}(T)=E_\mathrm{c}(0)+\Delta E_\mathrm{c}(T),
\end{align} 
is straightforward in the worldline picture by using the relation
(\ref{Int-4}). The finite-temperature correction is purely driven by the
worldlines with nonzero winding number. As the winding-number sum does not
take direct influence on the worldline averaging, the complicated
geometry-dependent part of the calculation remains the same for zero or finite
temperature. This disentangles the technical complications arising from
geometry on the one hand and temperature on the other hand in a convenient
fashion. The same statement holds for the Casimir force
$F_\mathrm{c}(T)=
F_\mathrm{c}(0)+\Delta F_\mathrm{c}(T)$.

\subsection{Parallel plates}

In order to demonstrate the simplicity of the worldline method, let us
calculate the well-known thermal contribution to the Casimir effect for parallel
plates. In the following, we use the dimensionless parameter
\begin{equation}
  \xi\equiv a T, \label{eq:1}
\end{equation}
which distinguishes between the high-temperature $\xi\gg 1$ and
low-temperature $\xi\ll 1$ parameter region.

Evaluating the general worldline formula for the Casimir energy \Eqref{Int-3}
using the parallel-plates $\Theta$ functional of \Eqref{T0-pp-1} results in
($D>2$)
\begin{align}\label{FT-pp-1}
 \frac{\Delta E_\mathrm{c}^\parallel(\xi)}{E_\mathrm{c}^\parallel(0)} =& 
\frac{\Gamma\left(\frac{D-1}{2}\right)\sqrt{\pi} \zeta(D-1)
\left(2 \xi\right)^{D-1}}{\Gamma(D/2)\zeta (D)} - \left(2
\xi\right)^{D}\notag
\\
+& \left\langle
\sum_{n=1}^\infty\frac{\lambda^D}{\Gamma(D/2)\zeta (D)}\right.
\\
&\left.\times\left[ E_{1-\frac{D}{2}}\left(\frac{\lambda^2 n^2}{4
\xi^2}\right)-E_{\frac{3}{2}-\frac{D}{2}}\left(\frac{\lambda^2 n^2}{4
\xi^2}\right) \right]\right\rangle,\notag
\end{align}
where the exponential integral function $E_n(z)$ is given by
\begin{align}\label{FT-pp-2}
E_{n}\left(z\right)\equiv \int_1^\infty \frac{e^{-zt}}{t^n} \ \mathrm{d}t,
\end{align}
and $\lambda$ again denotes the maximum extent of the worldline in the
direction orthogonal to the plates. In the low-temperature limit,
$2\xi\ll\sqrt{\langle\lambda^2\rangle}=\pi/\sqrt{3}$, the exponential integral
functions vanish exponentially and can be neglected. We then obtain the
small-temperature correction to $E_\mathrm{c}^\parallel(0)$ in $D$ dimensions
fully analytically:
\begin{align}\label{FT-pp-3}
\frac{\Delta E_\mathrm{c}^\parallel(\xi\rightarrow 0)}{E_\mathrm{c} ^\parallel(0)} &=
\frac{\Gamma\left(\frac{D-1}{2}\right)\sqrt{\pi} \zeta(D-1)
 \left(2 \xi\right)^{D-1}}{\Gamma(D/2)\zeta (D)} - \left(2
\xi\right)^{D}.
\end{align}
The term $\left(2 \xi\right)^{D}$ agrees with the standard textbook result
\cite{Milton:2001yy}. It dominates the thermal correction to the Casimir
force, yielding a comparatively suppressed power law dependence on the
temperature, $\Delta F_{\mathrm c}(T) \sim T^D$ for small $T$. This is an
immediate consequence of the gap in the relevant part of the fluctuation
spectrum in this closed geometry. This term can also be understood as an
excluded-volume effect: the volume in between the plates cannot be thermally
populated by photons at low temperature due to the spectral gap.

{Incidentally,} the leading contribution to the energy $\sim \xi^{D-1}$
is much less known. It does not contribute to the Casimir force, since it is
independent of $a$ when multiplied by the normalization prefactor
$E_\mathrm{c} ^\parallel(0)$. As we will see in section
\ref{sec:ft-inclinedPlates}, $a$-independent terms in the energy {should
  not be viewed as mere calculational artefacts but can also contribute to
  observables such as} the Casimir torque. {To the best of our
  knowledge, Eq.~(\ref{FT-pp-3}) represents} the first exact analytic formula
for this leading small temperature correction to the free energy.
 
\begin{figure}[t]
\begin{center}
\includegraphics[width=0.95\linewidth]{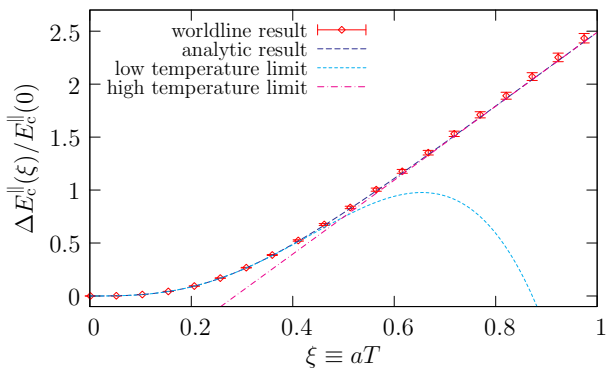}
\end{center}
\caption{Parallel plates: temperature dependence of the thermal contribution
  to the Casimir energy $\Delta
  E_\mathrm{c}^\parallel(\xi)/E_\mathrm{c}^\parallel(0)$ normalized to the
  zero-temperature result in $D=4$-dimensional spacetime versus the
  dimensionless temperature variable $\xi=a T$. The worldline result for 1000
  worldlines with $2\times10^6$ points each is plotted together with the analytic
  expressions (\ref{FT-pp-3})-(\ref{FT-pp-5}).  } \label{FT-pp-6}
\end{figure}

The high-temperature limit of (\ref{FT-pp-1}) can be obtained by a Poisson
resummation of the winding-number sum (which is identical to returning to
Matsubara frequency space). Our  result agrees with \cite{Milton:2001yy} and reads:
\begin{align}\label{FT-pp-4}
\frac{\Delta E_\mathrm{c}^\parallel(\xi\rightarrow
\infty)}{E_\mathrm{c}^\parallel(0)} =-1+\frac{2  \Gamma\left(\frac{D-1}{2}\right) \zeta(D-1) 
\sqrt{\pi}}{\Gamma(D/2) \zeta(D)}\xi.
\end{align}
For arbitrary $\xi$ and $D>2$, \Eqref{FT-pp-1} can be evaluated
numerically. Figure \ref{FT-pp-6} shows the worldline result together with the
analytic asymptotics (\ref{FT-pp-3})-(\ref{FT-pp-4}) and the known exact
analytic formula for $D=4$, see e.g. \cite{Feinberg:1999ys}:
\begin{align}\label{FT-pp-5}
&\frac{\Delta E_\mathrm{c}^\parallel(\xi)}{E_\mathrm{c}^\parallel(0)}=\notag
\\
&-1+\frac{90 \xi}{\pi^3}
\sum_{n=1}^\infty\frac{\coth(2n\pi \xi)+2 n \pi \xi\
\text{csch}^2(2 n \pi \xi)}{ n^3 }.
\end{align}
Note that the result obtained in \cite{Feinberg:1999ys} for the
electromagnetic field is twice as large as the result for the scalar field
(\ref{FT-pp-5}).

Of course, taking the derivative of (\ref{FT-pp-1}), (\ref{FT-pp-3}) or
(\ref{FT-pp-4}) with respect to $a$ also gives immediate access to the thermal
corrections $\Delta F_\mathrm{c}(T)$ to the Casimir force. For instance, the
low-temperature limit results in
\begin{align}
\Delta F_{\mathrm c}(aT\ll 1) &= - \frac{\partial }{\partial a} \Delta E^\|_{\text
    c}\Big|_{aT\ll 1} \notag
\\
&= - \frac{\Gamma(D/2) \zeta(D) A}{\pi^{D/2}}\, T^D, \label{eq:2}
\end{align}
again revealing the power-law suppressed temperature dependence which is
characteristic for a closed geometry. Both magnitude and sign of the thermal
force correction can be understood as an excluded-volume effect: as the
temperature is small compared to the spectral gap, thermal modes in-between the
plates cannot be excited. Hence, the thermal Stefan-Boltzmann energy density
outside the plates is not balanced by a thermal contribution inside. Thermal
effects therefore enhance the attractive force between the plates. 

Let us finally remark that the comparison between the small-temperature limit
of (\ref{FT-pp-5}) (calculated with the help of the Poisson summation) and our
analytic formula (\ref{FT-pp-3}) closes a gap in the literature.  With this
comparison, we can find the exact value of the integral occurring in the
prefactor of the leading low-temperature term in the energy,
\begin{align}\label{FT-pp-odd-Ref}
\int_0^\infty \mathrm{d} x \frac{1}{2 x^4}\left[-2+x(\coth x + x\text{
    csch}^2x)\right]=\frac{\zeta(3)}{2 \pi^2}, 
\end{align}
{numerically corresponding to $\approx 0.060897$. This result has
been observed numerically} in the sum over odd reflection
contributions to the parallel-plates Casimir energy in the optical
approach to the Casimir effect \cite{Scardicchio:2005di}.

\subsection{Inclined Plates}\label{sec:ft-inclinedPlates}

Whereas the inclined-plate geometry is much more difficult to deal with than
the parallel-plate case when using standard methods, there is comparatively
little difference in the worldline language. Inserting the inclined-plates
$\Theta$ functional (\ref{T0-ip-2}) with $\gamma_\mathrm{m}(x)$ as in
Eq.~(\ref{T0-ip-5b}) into the general worldline formula~(\ref{Int-3}) yields
\begin{align}\label{FT-ip-1}
E_\mathrm{c}^{\mathrm{i.p.},\varphi}(\xi)=E_\mathrm{c}^{\mathrm{i.p.},\varphi}+\Delta
E_\mathrm{c}^{\mathrm{i.p.},\varphi}(\xi),
\end{align}
where
\begin{widetext}
\begin{align}\label{FT-ip-2}
  \Delta E_\mathrm{c}^{\mathrm{i.p.},\varphi}(\xi)
  =-\frac{L_{y}\csc(\varphi)}{(4\pi)^{D/2} a^{D-2}} 
  &
  \left(\left(2 \xi\right)^{D-2}\zeta (D-2)
  \Gamma\left(\frac{D-2}{2}\right)
  \left\langle\int_{\gamma_{x_\mathrm{min}}}^{\gamma_{x_\mathrm{max}}}\!\!\!\!
  \mathrm{d}x
  \ \gamma_\mathrm{m}(x)\right\rangle
  - \zeta (D-1) \left(2 \xi\right)^{D-1}
  \Gamma\left(\frac{D-1}{2}\right) \sqrt{\pi} \right.\notag 
  \\
  &\left. 
  \,\,+ \left\langle \sum_{n=1}^\infty\int_{\gamma_{x_\mathrm{min}}}^{\gamma_\mathrm{xmax}} \mathrm{d}x \
  \gamma_\mathrm{m}^{D-1}(x)\times
  \ \ \ \left[E_{\frac{3}{2}-\frac{D}{2}}\left(\frac{\gamma_\mathrm{m}^2(x)
      n^2}{4 \xi^2}\right)-E_{2-\frac{D}{2}}\left(\frac{\gamma_\mathrm{m}^2(x) n^2 }{4
      \xi^2}\right)\right]\right\rangle\right) 
\end{align}
\end{widetext}
is the thermal contribution to the energy.  Here and in the following, we
confine ourselves to spacetime dimensions $D>3$ where all expressions exhibit
well-controlled convergence. In the low-temperature limit, the exponential
integral functions can be neglected as long as $\gamma_\mathrm{m}(x)\neq0$ for
all $x$.  This is certainly the case for $\varphi \neq 0$, but not necessarily
for $\varphi=0$. The latter case is again identical to the semi-infinite plate
parallel to a infinite one, and is being considered separately in the next
section and also in the appendix.  For $\varphi \neq 0$, the low-temperature
limit is then given by the first two terms (first line) of
Eq.~(\ref{FT-ip-2}). Note that the first $\xi^{D-2}$ term does not contribute
to the Casimir force, since it is an $a$-independent contribution to
$E_\mathrm{c}$ if read together with the normalization prefactor. From the
second term, we obtain the low-temperature thermal correction to the Casimir
force upon differentiation with respect to $a$,
\begin{equation}
\Delta F^{\mathrm{i.p.},\varphi\neq0} = -L_y \csc(\varphi)\,
\frac{\Gamma\left(\frac{D-1}{2}\right) \zeta(D-1)}{2 \pi^{(D-1)/2}} T^{D-1}.
\label{eq:3}
\end{equation}
The temperature dependence differs from the parallel-plates case by one power
of $T$, implying a significantly stronger temperature dependence at small
temperatures. This is a direct consequence of the fact that we are dealing
here with an open geometry. We emphasize that the result has been obtained
fully analytically. In $D=4$ and $\varphi=\pi/2$, our result agrees with the
perpendicular-plates study of \cite{Gies:2008zz} where this nontrivial
interplay between temperature and geometry has been demonstrated for the first
time.  As shown therein, the thermal correction for this open geometry at
experimentally-relevant large separations can be an order of magnitude larger
than for a closed geometry. 

Whereas the $a$-independent first term of Eq.~(\ref{FT-ip-2}) does not
contribute to the force, both terms in the first line of Eq.~(\ref{FT-ip-2})
contribute to the low-temperature limit of the Casimir torque. The thermal
contribution to the torque is
\begin{align}\label{FT-ip-3}
\Delta D_\mathrm{c}^\mathrm{i.p.,\varphi}(\xi)=\frac{\mathrm{d}\Delta
  E_\mathrm{c}^\mathrm{i.p.,\varphi}(\xi)}{\mathrm{d}\varphi},
\end{align}
which at low temperature reads
\begin{align}\label{FT-ip5}
\Delta  D_\mathrm{c}^{\mathrm{i.p.},\varphi}&(\xi\rightarrow 0)
=-\frac{L_{y}\cos(\varphi)}{(4\pi)^{D/2} \sin^2(\varphi)}\notag
\\
&\!\!\!\!\!\!\!\!\! \times\left(\left(2
T\right)^{D-2}\zeta (D-2)
\Gamma\left(\frac{D-2}{2}\right)\right.\notag
\\
&\left. \ \ \times\left[\langle
\gamma_{x_\mathrm{min}}(\varphi)\lambda_x \rangle-{\tan(\varphi)\langle
\gamma_{x_\mathrm{min}}'(\varphi)\lambda_x \rangle}\right] \right. \notag
\\
&\left.\ \ +a\zeta (D-1) \left(2 T\right)^{D-1}
\Gamma\left(\frac{D-1}{2}\right) \sqrt{\pi}\right),
\end{align}
where
$\left\langle\int_{\gamma_{x_\mathrm{min}}}^{\gamma_{x_\mathrm{max}}}x\mathrm{d}x\right\rangle=0$
has been used, and we have introduced
$\lambda_x=\gamma_{x_\mathrm{max}}-\gamma_{x_\mathrm{min}}$.  This expression
depends on only one nontrivial worldline average. In limiting cases, this
average can be given analytically, as it reduces to the case described by
Eq.~(\ref{T0-pp-4}): we find
$-\langle \gamma_{x_\mathrm{min}}(\varphi\to 0)\lambda_x \rangle =
\langle\lambda^2_x \rangle/2=\pi^2/6$ and $-\langle
\gamma_{x_\mathrm{min}}(\varphi\to \pi/2)\lambda_x \rangle = \langle\lambda_x
\rangle^2/2=\pi/2$. For arbitrary $\varphi$, this average can be well
approximated by
\begin{align}\label{FT-ip6}
\langle \bgam_{x_\mathrm{min}}(\varphi)\lambda_x \rangle\approx
-\frac{\pi}{2}\sin^2(\varphi)-\frac{\pi^2}{6}\cos^2(\varphi),
\end{align}
as we will explain in the following. With  Eq.~(\ref{T0-ip-3}), we can write 
\begin{align}\label{ft-ip-6-1}
\bgam_{x_{\mathrm{min}}}(\varphi)
  \equiv \bgam_{{x}}(\hat{t})\cos(\varphi)
  +\bgam_{z}(\hat{t})\sin(\varphi),
\end{align}
where $\hat{t}$ denotes the value of $t$ {that satisfies the minimum
  condition} in Eq.~(\ref{T0-ip-3}).  Together with
\begin{align}\label{ft-ip-6-2}
\bgam_{z_{\mathrm{min}}}(\varphi)
  \equiv -\bgam_{x}(\hat{t})\sin(\varphi)+\bgam_{{z}}(\hat{t})\cos(\varphi),
\end{align}
we can interpret
$(\bgam_{x_\mathrm{min}}(\varphi),\bgam_{z_\mathrm{min}}(\varphi))$ as
the coordinates of the point $(\bgam_{{x}}(\hat{t}),
\bgam_{z}(\hat{t}))$  {in the $\varphi$-rotated system.}

\begin{figure}[t]
\begin{center}
\scalebox{0.7}{
\begin{picture}(0,0)%
\includegraphics{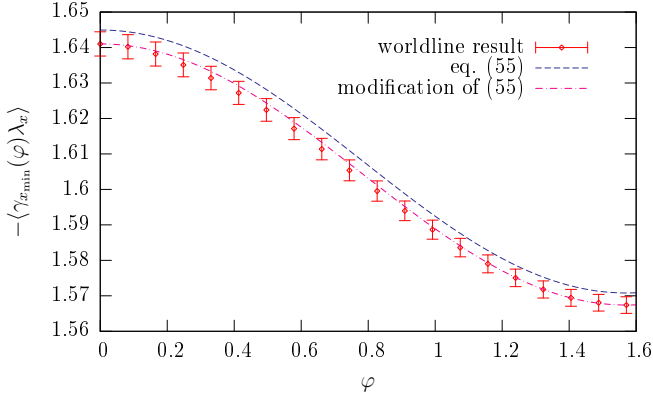}%
\end{picture}%
\begingroup
\setlength{\unitlength}{0.0200bp}%
\begin{picture}(18000,10800)(0,0)%
\large
\put(2475,1650){\makebox(0,0)[r]{\strut{} 1.56}}%
\put(2475,2606){\makebox(0,0)[r]{\strut{} 1.57}}%
\put(2475,3561){\makebox(0,0)[r]{\strut{} 1.58}}%
\put(2475,4517){\makebox(0,0)[r]{\strut{} 1.59}}%
\put(2475,5472){\makebox(0,0)[r]{\strut{} 1.6}}%
\put(2475,6428){\makebox(0,0)[r]{\strut{} 1.61}}%
\put(2475,7383){\makebox(0,0)[r]{\strut{} 1.62}}%
\put(2475,8339){\makebox(0,0)[r]{\strut{} 1.63}}%
\put(2475,9294){\makebox(0,0)[r]{\strut{} 1.64}}%
\put(2475,10250){\makebox(0,0)[r]{\strut{} 1.65}}%
\put(2750,1100){\makebox(0,0){\strut{} 0}}%
\put(4553,1100){\makebox(0,0){\strut{} 0.2}}%
\put(6356,1100){\makebox(0,0){\strut{} 0.4}}%
\put(8159,1100){\makebox(0,0){\strut{} 0.6}}%
\put(9963,1100){\makebox(0,0){\strut{} 0.8}}%
\put(11766,1100){\makebox(0,0){\strut{} 1}}%
\put(13569,1100){\makebox(0,0){\strut{} 1.2}}%
\put(15372,1100){\makebox(0,0){\strut{} 1.4}}%
\put(17175,1100){\makebox(0,0){\strut{} 1.6}}%
\put(550,5950){\rotatebox{90}{\makebox(0,0){\strut{}$-\langle   \gamma_{x_\mathrm{min}}(\varphi) \lambda_x\rangle$}}}%
\put(9962,275){\makebox(0,0){\strut{}$\varphi$}}%
\put(14195,9294){\makebox(0,0)[r]{\strut{}worldline result}}%
\put(14195,8744){\makebox(0,0)[r]{\strut{}eq. (\ref{FT-ip6})}}%
\put(14195,8194){\makebox(0,0)[r]{\strut{}modification of (\ref{FT-ip6})}}%
\normalsize
\end{picture}%
\endgroup
}
\end{center}
\caption{Angle dependence of the worldline average occurring in the Casimir
  torque (\ref{FT-ip5}): $\langle
  \gamma_{x_{\text{min}}}(\varphi)\lambda_x\rangle$. The numerical result is
  compared to the estimate (\ref{FT-ip6}) (dashed line). The worldline result
  shows a small systematic error due to the finite discretization. We can
  include the systematic error into (\ref{FT-ip6}) by taking worldline
  {estimates for the} boundary values at $\varphi=0$ and $\varphi=\pi/2$
  (i.e. $1.567 \sin^2(\varphi)+1.641 \cos^2(\varphi)$), respectively, which
  are smaller than $\pi/2$ and $\pi^2/6$. This yields the dot-dashed curve.
  The worldline result has been obtained from $5 \ 10^4$ loops with $10^6$ ppl
  each.  } \label{FT-ip8}
\end{figure}

Since the $\bgam_x$ and $\bgam_z$ coordinates of each loop are generated
independently of each other, $\bgam_{z}(\hat{t})$ and $\lambda_x$ are not
correlated. We therefore obtain
\begin{align}\label{ft-ip-6-3}
\langle \lambda_x \bgam_{x_\mathrm{min}}(\varphi)\rangle=\langle\lambda_x\bgam_{x}(\hat{t})\rangle\cos(\varphi)+\langle\lambda_x\rangle\langle\bgam_{z}(\hat{t})\rangle\sin(\varphi).
\end{align}
By symmetry, the average
 $\langle\bgam_{z_\mathrm{min}}(\varphi)\rangle$ vanishes, and we get from
 Eq.~(\ref{ft-ip-6-2})
\begin{align}\label{ft-ip-6-4}
\langle\bgam_{x}(\hat{t})\rangle \sin(\varphi)=\langle\bgam_{{z}}(\hat{t})\rangle \cos(\varphi).
\end{align}
On the other hand, $ \langle \bgam_{x_{\mathrm{min}}}(\varphi) \rangle=-\langle
\lambda_x\rangle/2$.  Substituting Eq.~(\ref{ft-ip-6-4}) into the average of
Eq.~(\ref{ft-ip-6-1}) leads to $\langle
\bgam_{z}(\hat{t})\rangle=-\sin(\varphi)\langle\lambda_x\rangle/2$ and
$\langle \bgam_{x}(\hat{t})\rangle=-\cos(\varphi)\langle\lambda_x\rangle/2$,
{such that the desired} Eq.~(\ref{FT-ip6}) can be motivated by
Eq.~(\ref{ft-ip-6-3}). {We would like to stress} that only the second
term in Eq.~(\ref{FT-ip6}) has been estimated {with the constraint
  imposed by the exactly known result for $\varphi\to0$, see above. The first
  term} is exact and dictates the behavior of the perpendicular-plates limit.
This result is compared to the numerically obtained data in Fig.~\ref{FT-ip8}.

For the v-loop algorithm \cite{Gies:2006cq} used here to generate the
loops, the expectation value of the maximal extent $\lambda_x$ is
systematically smaller. This error is about the average {spacing}
$\langle|\bgam_{x}(t_{i+1})-\bgam_{x}(t_{i})|\rangle\approx\sqrt{1/2 N}$, see
Eq.~(\ref{intro-10}). As a consequence, the systematic error of
$\langle\lambda_x^2\rangle/2$ and $\langle\lambda_x\rangle^2/2$ is about
$\langle\lambda_x\rangle\sqrt{1/2 N}\approx 1.3\times10^{-3}$ at $N=10^6$. We
observe a good agreement of the data with Eq.(\ref{FT-ip6}) at $\varphi=0$ and
$\varphi=\pi/2$ if the systematic error is taken into account. However, the
agreement is actually perfect for all $\varphi$ when using worldline
{estimates for the prefactors in Eq.~(\ref{FT-ip6})} instead of $\pi/2$
and $\pi^2/6$, {see the modified curve in Fig.~\ref{FT-ip8}.} This shows
that Eq.~(\ref{FT-ip6}) will well fit the data in the {continuum limit}
$N\rightarrow \infty$. 

Let us return to the calculation of the Casimir torque. In the vicinity
of the perpendicular-plates configuration, $\varphi=\pi/2 {-} \delta \varphi$,
we can now obtain an expression to first order in $\delta
\varphi$:
\begin{align}\label{FT-ip9}
\Delta & D_\mathrm{c}^{\mathrm{i.p.},\varphi=\pi/2 {-} \delta \varphi}(\xi\rightarrow 0)
={-}\frac{L_{y}\delta\varphi}{(4\pi)^{D/2}}\, \left(2
T\right)^{D-2}\notag
\\
& \ \ \ \ \ \times\Big(-\zeta (D-2)
\Gamma\left(\frac{D-2}{2}\right)\left[\frac{\pi}{2}-{\left\langle
\gamma_{x_\mathrm{min}}''\left(\frac{\pi}{2}\right)\lambda_x \right\rangle}\right] \notag
\\
& \ \ \ \ \ \ \ \quad +2\,aT\, \zeta (D-1) 
\Gamma\left(\frac{D-1}{2}\right) \sqrt{\pi}\Big).
\end{align}
Here we have used $\langle \gamma_{x_\mathrm{min}}'(\pi/2)\lambda_x
\rangle=0$. Apart from $\langle
\gamma_{x_\mathrm{min}}''(\pi/2)\lambda_x\rangle$, Eq. (\ref{FT-ip9}) {is
  an analytical expression}. Using (\ref{FT-ip6}), we obtain $-\langle
\gamma_{x_\mathrm{min}}''(\pi/2)\lambda_x\rangle\approx \pi^2/3-\pi\approx
0.148$, which is about ten percent of the {dominating analytical} term
{in square brackets} $\sim \pi/2$. {We observe that the first term,
  which dominates in the limit $aT\to 0$, gives a contribution to the torque
  which drives the system away from the perpendicular-plates case
$\varphi=\pi/2$. Zero- and finite-temperature contributions thus have the same
sign. The fact that $\varphi=\pi/2$ is a repulsive fixed point is also in
agreement with naive expectations.}

{For $D=4$, Eq. (\ref{FT-ip9}) reads
\begin{align}\label{FT-ip9b}
\frac{\Delta  D_\mathrm{c}^{\mathrm{i.p.},\varphi=\pi/2 - \delta \varphi}(\xi\rightarrow 0)}{L_y}
=\delta\varphi T^2\left(0.0716-0.0957 \xi\right),
\end{align}
which should be compared with the first-order term arising from the $T=0$
contribution Eq.~(\ref{T0-ip-torque-2}), which reads
$D_\mathrm{c}^\mathrm{i.p.,\varphi=\pi/2 - \delta \varphi}
/L_y\approx0.00329\delta \varphi/a^2$.
Thus, for $D=4$ we obtain to first order in $\delta\varphi$
\begin{align}\label{FT-ip9c}
\frac{\Delta  D_\mathrm{c}^{\mathrm{i.p.},\varphi=\pi/2 - \delta
    \varphi}(\xi\rightarrow 0)}{D_\mathrm{c}^{\mathrm{i.p.},\varphi=\pi/2 -
    \delta \varphi}(0)} \approx \xi^2\left(21.8-29.1 \xi\right).
\end{align}
} {In the validity regime of the low-temperature expansion, $\xi=aT\ll
  1$, the positive first term is always dominant, hence perpendicular-plates
  case remains a repulsive fixed point. Most importantly, we would like to
  stress that the quadratic dependence of the torque on the temperature $\sim
  T^2$ ($\sim T^{D-2}$ in the general case) for the inclined-plates
  configuration represents the strongest temperature dependence of all
  observables discussed in this article. }

\begin{figure}[t]
\begin{center}
\scalebox{0.7}{
\begin{picture}(0,0)%
\includegraphics{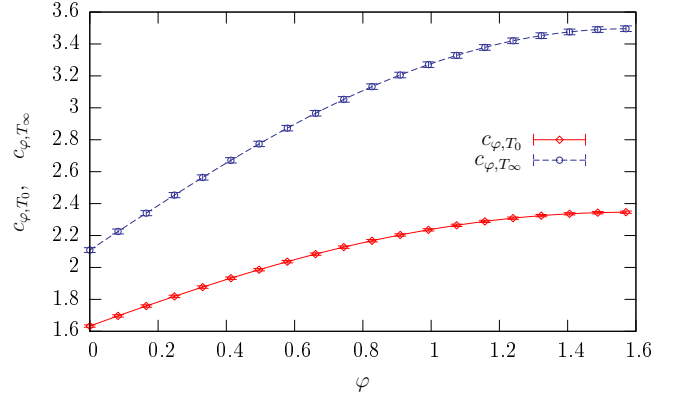}%
\end{picture}%
\begingroup
\setlength{\unitlength}{0.0200bp}%
\begin{picture}(18000,10800)(0,0)%
\large
\put(2200,1650){\makebox(0,0)[r]{\strut{} 1.6}}%
\put(2200,2510){\makebox(0,0)[r]{\strut{} 1.8}}%
\put(2200,3370){\makebox(0,0)[r]{\strut{} 2}}%
\put(2200,4230){\makebox(0,0)[r]{\strut{} 2.2}}%
\put(2200,5090){\makebox(0,0)[r]{\strut{} 2.4}}%
\put(2200,5950){\makebox(0,0)[r]{\strut{} 2.6}}%
\put(2200,6810){\makebox(0,0)[r]{\strut{} 2.8}}%
\put(2200,7670){\makebox(0,0)[r]{\strut{} 3}}%
\put(2200,8530){\makebox(0,0)[r]{\strut{} 3.2}}%
\put(2200,9390){\makebox(0,0)[r]{\strut{} 3.4}}%
\put(2200,10250){\makebox(0,0)[r]{\strut{} 3.6}}%
\put(2475,1100){\makebox(0,0){\strut{} 0}}%
\put(4313,1100){\makebox(0,0){\strut{} 0.2}}%
\put(6150,1100){\makebox(0,0){\strut{} 0.4}}%
\put(7988,1100){\makebox(0,0){\strut{} 0.6}}%
\put(9825,1100){\makebox(0,0){\strut{} 0.8}}%
\put(11663,1100){\makebox(0,0){\strut{} 1}}%
\put(13500,1100){\makebox(0,0){\strut{} 1.2}}%
\put(15337,1100){\makebox(0,0){\strut{} 1.4}}%
\put(17175,1100){\makebox(0,0){\strut{} 1.6}}%
\put(550,5950){\rotatebox{90}{\makebox(0,0){\strut{}$\const_{\varphi,T_0}, \ \ \ \const_{\varphi,T_\infty}$}}}%
\put(9825,275){\makebox(0,0){\strut{}$\varphi$}}%
\put(14144,6810){\makebox(0,0)[r]{\strut{}$\const_{\varphi,T_0}$}}%
\put(14144,6260){\makebox(0,0)[r]{\strut{}$\const_{\varphi,T_\infty}$}}%
\normalsize
\end{picture}%
\endgroup
}
\end{center}
\caption{{Integrals appearing in Eqs. (\ref{FT-ip-2}) and (\ref{FT-ip-10}) for
    $D=4$. We have defined
    $\const_{\varphi,T_0}\equiv\left\langle\int_{\gamma_{x_\mathrm{min}}}^{\gamma_{x_\mathrm{max}}}
    \mathrm{d}x \ \gamma_\mathrm{m}(x)\right\rangle$ and
    $\const_{\varphi,T_\infty}\equiv\left\langle\int_{\gamma_{x_\mathrm{min}}}^{\gamma_{x_\mathrm{max}}}
    \mathrm{d}x \ \gamma_\mathrm{m}^2(x)\right\rangle$; see also
    Eqs.~(\ref{int-sum-6}), (\ref{int-sum-7}) and (\ref{int-sum-9}).
    Employing Eq.~(\ref{T0-pp-4}), we can evaluate
    $\const_{\varphi=0,T_0}=\zeta(2)\approx 1.645$ and
    $\const_{\varphi=0,T_\infty}=\sqrt{\pi}\zeta(3)\approx 2.131$
    analytically. We have used $10^4$ worldlines with $10^6$ ppl
    each.}}\label{ym-integrals}
\end{figure}

The high-temperature limit of Eq.~(\ref{FT-ip-2}) can again be obtained by Poisson
summation. The result is:
\begin{align}\label{FT-ip-10}
\Delta E_\mathrm{c}^{\mathrm{i.p.},\varphi}(\xi & \rightarrow 
\infty)=-E_\mathrm{c}^{\mathrm{i.p.},\varphi}(0)\notag
\\
&-\frac{L_{y}2\sqrt{\pi} \left\langle\int_{\gamma_{x_\mathrm{min}}}^{\gamma_{x_\mathrm{max}}}
\gamma_\mathrm{m}^{D-2}(x)  \mathrm{d}x\right\rangle}{(4\pi)^{D/2}
a^{D-2}(D-3)(D-2)\ \sin(\varphi)}\xi .
\end{align}
The remaining worldline average in this expression yields some positive finite
number. Irrespective of its precise value for a given angle $\varphi$ and $D$
{(the precise value of the integral for a specific $\varphi$ {can be read
    off, for instance, from }
  Fig. \ref{ym-integrals} for either $D=3$ or $D=4$ and Fig. \ref{T0-ip-7}
  for $D=5$)}, we stress that we observe the same linear dependence on
temperature $\xi=aT$ as in the parallel plate case. This is nothing but the
familiar phenomenon of the dominance of the zeroth Matsubara mode at high
temperatures, implying dimensional reduction, as discussed above. This
mechanism is obviously geometry independent. Also the Casimir force remains
attractive also for high temperatures.

\subsection{Semi-infinite plate parallel to an infinite plate}

A particularly interesting example for the geometry-temperature interplay is
given by the semi-infinite plate parallel to the infinite plate (1si
configuration).  In this case, the angle of inclination $\varphi$ in
Fig. \ref{T0-ip-4} is zero. Analogously to Eq.~(\ref{T0-ip-4}), the
finite-temperature Casimir energy can be decomposed as 
\begin{align}\label{FT-si-1}
E_\mathrm{c}^\mathrm{1si}(\xi)
=E_\mathrm{c}^\mathrm{1si,edge}(\xi)+E_\mathrm{c}^\mathrm{1si,\parallel}(\xi),
\end{align}
where
$E_\mathrm{c}^\mathrm{1si,\parallel}(T)=E_\mathrm{c}^\mathrm{1si,\parallel}(0)+\Delta
E_\mathrm{c}^\mathrm{1si,\parallel}(T)$ correspond to the standard
parallel-plate formulas as given in Eqs.~(\ref{T0-pp-3}) and (\ref{FT-pp-1}),
with $A$ being now the surface of the semi-infinite plate. Approaching the 1si
limit of $\Delta E_\mathrm{c}^\mathrm{1si,\parallel}(T)$ from the
inclined-plates configuration in the limit $\varphi\to 0$ is again a delicate
issue, as the proper order of limits $\varphi\to 0$ and $L_z\to \infty$ has to
be accounted for, see Sect.~\ref{sec:incl-plat-varph}. As the analysis is
technically involved (but the outcome obvious), we defer it to the appendix.
Let us here concentrate on the temperature-dependent edge contribution $\Delta
E_\mathrm{c}^\mathrm{1si,edge}(T)
=E_\mathrm{c}^\mathrm{1si,edge}(T)-E_\mathrm{c}^\mathrm{1si,edge}(0)$. We set
$\varphi=0$ in Eq.~(\ref{T0-ip-2}) and evaluate Eq.~(\ref{Int-3}). The result
is (here and in the following, we confine ourselves to $D>3$):
\begin{widetext}
  \begin{align}\label{FT-si-2}
    \Delta E_\mathrm{c}^\mathrm{1si,edge}(\xi)=-\frac{L_{y}}{(4\pi)^{D/2}
      a^{D-2}}&\left[ \left(2
      \xi\right)^{D-2}\zeta (D-2)
      \Gamma\left(\frac{D-2}{2}\right)
      \left\langle\int_{\gamma_{x_\mathrm{min}}}^{\gamma_{x_\mathrm{max}}}
      \mathrm{d}x\  \gamma_{z_{\mathrm{max}}}(x)\notag \right\rangle\right.
      \\
      &\left.-
      \sum_{n=1}^\infty\left\langle\int_{\gamma_{x_\mathrm{min}}}^{\gamma_{x_\mathrm{max}}}
      \mathrm{d}x\ (x-\gamma_{x_\mathrm{min}})^{D-2} \gamma_{z_{\mathrm{max}}}(x)
      E_{2-\frac{D}{2}}\left(\frac{(x-\gamma_{x_\mathrm{min}})^2 n^2 }{4
        \xi^2}\right)\right\rangle\right].
\end{align}
\end{widetext}
Note that the first term, being the main contribution to the Casimir energy at
small $T$, does not contribute to the Casimir force since it is $a$
independent. Contrary to the case with $\varphi\neq0$, the exponential
integral functions cannot be neglected in the low-temperature limit, since the
argument of $E_n(z)$ becomes zero at the lower bound of the integral for any
$\xi>0$. This results in a correction $\sim \xi^{D-1+\alpha}$, with $\alpha>0$,
to the low-temperature limit of the first term.

Here, however, we concentrate on the last term, as it gives rise to the
thermal correction of the Casimir force. In the low-temperature limit, we find
\begin{align}\label{FT-si-3}
  -&\Delta F_\mathrm{c}^\mathrm{1si,edge} (\xi)
  =\frac{2 L_{y}}{ (4\pi)^{D/2}a^{D-1}} 
\\
  &\sum_{n=1}^\infty\left\langle\int_{0}^{\lambda_x}
  \mathrm{d}x \ x^{D-2} \gamma_{z_{\mathrm{max}}}(\gamma_{x_\mathrm{min}}+x)
  \exp\left(-\frac{x^2 n^2 }{4
    \xi^2}\right)\right\rangle .\notag
\end{align}
For small $\xi$, the main contribution to the integral comes from its lower
bound as the exponential function rapidly decreases for large argument.  At
the lower bound, we can take the worldline average $\langle
\gamma_{z_{\mathrm{max}}}(\gamma_{x_\mathrm{min}}+x)\rangle$ first, yielding
a smooth function. We expand the latter in a power series,
\begin{figure}[t]
\begin{center}
\scalebox{0.7}{
%
%
%
\begin{picture}(0,0)%
\includegraphics{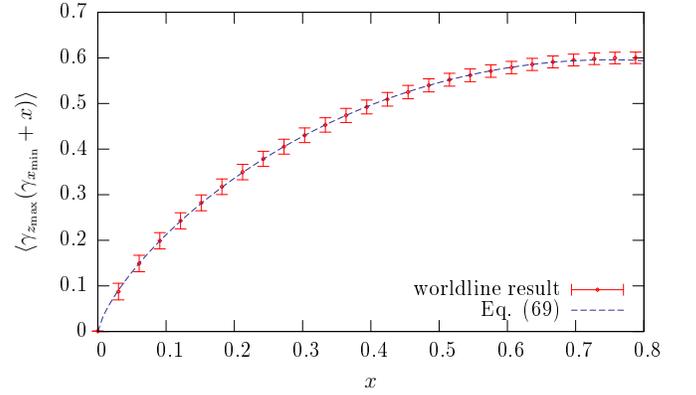}%
\end{picture}%
\begingroup
\setlength{\unitlength}{0.0200bp}%
\begin{picture}(18000,10800)(0,0)%
\large
\put(2200,1650){\makebox(0,0)[r]{\strut{} 0}}%
\put(2200,2879){\makebox(0,0)[r]{\strut{} 0.1}}%
\put(2200,4107){\makebox(0,0)[r]{\strut{} 0.2}}%
\put(2200,5336){\makebox(0,0)[r]{\strut{} 0.3}}%
\put(2200,6564){\makebox(0,0)[r]{\strut{} 0.4}}%
\put(2200,7793){\makebox(0,0)[r]{\strut{} 0.5}}%
\put(2200,9021){\makebox(0,0)[r]{\strut{} 0.6}}%
\put(2200,10250){\makebox(0,0)[r]{\strut{} 0.7}}%
\put(2475,1100){\makebox(0,0){\strut{} 0}}%
\put(4313,1100){\makebox(0,0){\strut{} 0.1}}%
\put(6150,1100){\makebox(0,0){\strut{} 0.2}}%
\put(7988,1100){\makebox(0,0){\strut{} 0.3}}%
\put(9825,1100){\makebox(0,0){\strut{} 0.4}}%
\put(11663,1100){\makebox(0,0){\strut{} 0.5}}%
\put(13500,1100){\makebox(0,0){\strut{} 0.6}}%
\put(15337,1100){\makebox(0,0){\strut{} 0.7}}%
\put(17175,1100){\makebox(0,0){\strut{} 0.8}}%
\put(550,5950){\rotatebox{90}{\makebox(0,0){\strut{}$\langle \gamma_{z_{\mathrm{max}}}(\gamma_{x_\mathrm{min}}+x)\rangle$}}}%
\put(9825,275){\makebox(0,0){\strut{}$x$}}%
\put(14950,2775){\makebox(0,0)[r]{\strut{} worldline result}}%
\put(14950,2225){\makebox(0,0)[r]{\strut{} Eq. (\ref{FT-si-8})}}%
\normalsize
\end{picture}%
\endgroup
}
\end{center}
\caption{Numerical result of the worldline average $\langle
  \gamma_{z_{\mathrm{max}}}(\gamma_{x_\mathrm{min}}+x)\rangle$ obtained by
  $5\ 10^4$ worldlines with $10^6$ points each compared to the fit function of
  Eq. (\ref{FT-si-8}) obtained on the interval $x=[0,0.7]$. The error bars 
  have been plotted ten times larger. The observed small-$x$ power law
  $x^{\alpha_1}$ with $\alpha_1 \simeq 0.74$ directly translates into a
  non-integer small-temperature behavior of the thermal edge contribution to
  the force, $\Delta F_{\text{c}}^{\text{1si,edge}}\sim T^{D-1+\alpha_1}$.}\label{FT-si-7}
\end{figure}
\begin{align}\label{FT-si-4}
\langle
\gamma_{z_{\mathrm{max}}}(\gamma_{x_\mathrm{min}}+x)\rangle=\sum_{n=0}^\infty
c_n x^{\alpha_n}, 
\end{align}
where the exponents $\alpha_n$ do not necessarily have to be integers.
Inserting (\ref{FT-si-4}) into (\ref{FT-si-3}) leads to
\begin{figure}[t]
\begin{center}
\scalebox{0.7}{
\begin{picture}(0,0)%
\includegraphics{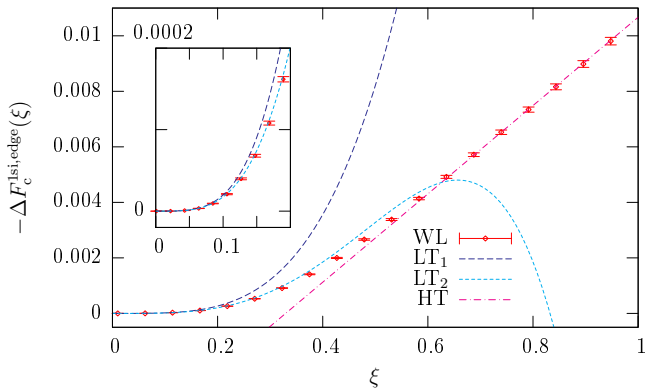}%
\end{picture}%
\begingroup
\setlength{\unitlength}{0.0200bp}%
\begin{picture}(18000,10800)(0,0)%
\large
\put(2750,2024){\makebox(0,0)[r]{\strut{} 0}}%
\put(2750,3520){\makebox(0,0)[r]{\strut{} 0.002}}%
\put(2750,5015){\makebox(0,0)[r]{\strut{} 0.004}}%
\put(2750,6511){\makebox(0,0)[r]{\strut{} 0.006}}%
\put(2750,8007){\makebox(0,0)[r]{\strut{} 0.008}}%
\put(2750,9502){\makebox(0,0)[r]{\strut{} 0.01}}%
\put(3025,1100){\makebox(0,0){\strut{} 0}}%
\put(5855,1100){\makebox(0,0){\strut{} 0.2}}%
\put(8685,1100){\makebox(0,0){\strut{} 0.4}}%
\put(11515,1100){\makebox(0,0){\strut{} 0.6}}%
\put(14345,1100){\makebox(0,0){\strut{} 0.8}}%
\put(17175,1100){\makebox(0,0){\strut{} 1}}%
\put(550,5950){\rotatebox{90}{\makebox(0,0){\strut{}$-\Delta F_\mathrm{c}^\mathrm{1si,edge}(\xi)$}}}%
\put(10100,275){\makebox(0,0){\strut{}$\xi$}}%
\put(12089,4043){\makebox(0,0)[r]{\strut{}WL}}%
\put(12089,3493){\makebox(0,0)[r]{\strut{}LT$_1$}}%
\put(12089,2943){\makebox(0,0)[r]{\strut{}LT$_2$}}%
\put(12089,2393){\makebox(0,0)[r]{\strut{}HT}}%
\put(3915,4779){\makebox(0,0)[r]{\strut{} 0}}%
\put(3915,6975){\makebox(0,0)[r]{\strut{} }}%
\put(5215,9570){\makebox(0,0)[r]{\strut{} 0.0002}}%
\put(4190,3790){\makebox(0,0){\strut{} 0}}%
\put(5096,3790){\makebox(0,0){\strut{} }}%
\put(6003,3790){\makebox(0,0){\strut{} 0.1}}%
\put(6909,3790){\makebox(0,0){\strut{} }}%
\put(7815,3790){\makebox(0,0){\strut{} }}%
\normalsize
\end{picture}%
\endgroup
}
\end{center}
\caption{Thermal contribution to the Casimir edge force in the 1si
  configuration, $-\Delta F_\mathrm{c}^\mathrm{1si,edge} (\xi)$, plotted for
  $a=1$ and $D=4$. WL: worldline result Eq.~(\ref{FT-si-3}) obtained using
  $1000$ loops with $10^6$ ppl each. LT$_1$, LT$_2$: leading and
  next-to-leading low-temperature corrections $0.1098\xi^{3.7423}$ and
  $0.1098\xi^{3.7423}-0.131881\xi^{4.7423}$, respectively, obtained from
  Eqs.~(\ref{FT-si-5}) and (\ref{FT-si-8}), using $5\times10^4$ loops $10^6$ ppl
  each. HT: high temperature limit obtained from Eq. (\ref{FT-si-9}), using
  $5\times10^4$ loops $10^6$ ppl each; a fit to the HT curve is provided by
  $-5.24062(\pm 0.0222) 10^{-3} + 1.591 (\pm 0.004138) 10^{-2} \xi $.  The
  inlay displays a magnified interval $\xi=[0,0.2]$.}\label{Fig-si-9}
\end{figure}
\begin{align}\label{FT-si-5}
-&\Delta F_\mathrm{c}^\mathrm{1si,edge}(\xi\rightarrow
0)=\frac{L_{y}}{(4\pi)^{D/2} a^{D-1}} \\
&\times\sum_{n=0}^\infty 
c_n \left(2 \xi\right)^{D+\alpha_n -1} \Gamma \left( \frac{D+\alpha_n
-1}{2}\right)\zeta(D+\alpha_n -1),\notag
\end{align}
where we have neglected exponentially suppressed contributions.  {For the
  lowest-order term,} we obtain $\langle
\gamma_{z_{\mathrm{max}}}(\gamma_{x_\mathrm{min}})\rangle\equiv c_0=0$, since
for a given worldline $(\bgam_x(t),\bgam_z(t))$ there exists a corresponding
worldline in the ensemble with $(\bgam_x(t),-\bgam_z(t))$.  {We conclude
  that} the coefficient of the $ T^{D-1}$ term vanishes.  We determine the
higher coefficients $c_n$ from computing $\langle
\gamma_{z_{\mathrm{max}}}(\gamma_{x_\mathrm{min}}+x)\rangle$ in the vicinity
of $x=0$ by worldline numerics.  Figure \ref{FT-si-7} depicts the form of
$\langle \gamma_{z_{\mathrm{max}}}(\gamma_{x_\mathrm{min}}+x)\rangle$ near the
lower bound $x=0$. A global fit to this function including two coefficients
$c_1,c_2$ is given by
\begin{align}\label{FT-si-8}
\langle \gamma_{z_{\mathrm{max}}}(\gamma_{x_\mathrm{min}}+x)\rangle
&\approx0.9132\left(x(1.500-x)\right)^{0.7423}\notag
\\&\approx 1.234 x^{0.7423}-0.6106 x^{1.7423},
\end{align}
where we have kept $\alpha_2-\alpha_1=1$ fixed. The resulting thermal
correction to the force is shown in Fig.~(\ref{Fig-si-9}) for $D=4$, where we
compare the full numerical solution with different orders of the expansion
\eqref{FT-si-8} and the high-temperature asymptotics, see below. As the
low-temperature asymptotics is directly related to the lowest nonvanishing
coefficient $\alpha_1$, we have also performed local fits to the function
$\langle \gamma_{z_{\mathrm{max}}}(\gamma_{x_\mathrm{min}}+x)\rangle$ in the
vicinity of $x=0$. Depending on the fit window, the leading exponent can grow
up to $\alpha_1\approx0.8$. (Of course, the fit window must be large enough to
avoid that the worldline discretization becomes visible; otherwise, the
exponent trivially but artificially approaches $\alpha_1\to 1$ as the
discretized worldline is a polygon on a microscopic scale). 

In any case, we conclude that the low-temperature regime of the 1si edge
effect is well described by a non-integer power law, $
\Delta F_{\text{c}}^{\text{1si,edge}}\sim T^{D-1+\alpha_1}\simeq T^{D-0.3}$,
where the fractional exponent arises from the geometry-temperature
interplay in this open geometry. Of course, our numerical analysis cannot
guarantee to 
yield the true asymptotic behavior in the limit $\xi\to 0$, but our data in
the low-temperature domain $0.01\lesssim \xi\lesssim 0.4$ is well described by
the non-integer scaling at next-to-leading order. 

Let us finally turn to the high-temperature limit of \Eqref{FT-si-2} which can
again be obtained by Poisson summation. The result for the edge energy reads
\begin{align}\label{FT-si-9} 
\Delta &E_\mathrm{c}^\mathrm{1si,edge}(\xi\rightarrow \infty)=-
E_\mathrm{c}^\mathrm{1si,edge}(0)\\
&-\frac{2 \sqrt{\pi } L_{y} \left\langle\int_{0}^{\lambda_{x}}
\mathrm{d} x \ x^{D-3}
\gamma_{z_{\mathrm{max}}}(x+\gamma_{x_\mathrm{min}})\right\rangle}{(4\pi)^{D/2}
a^{D-2} (D-3)}  \  \xi,\notag
\end{align}
where the worldline average is subject to numerical evaluation. The resulting
high-temperature limit of the Casimir force is shown in Fig. 
(\ref{FT-si-9}) for $D=4$. The high-temperature limit is again linear in $T$
in accordance with general dimensional-reduction arguments.

\section{Conclusions}
\label{sec:conclusions}

In this work, we have provided further numerical as well as analytical
evidence for the nontrivial interplay between geometry and temperature in the
Casimir effect. Whereas closed geometries such as the parallel-plates case
exhibit a comparatively strong suppression of thermal corrections at low
temperatures, open geometries such as the general inclined-plates geometry
reveal a more pronounced temperature dependence in this regime. The
terminology {\em open} and {\em closed} corresponds to the absence or presence
of a gap in the relevant part of the spectrum of fluctuations which gives rise
to the Casimir effect. In closed geometries, the spectral gap inhibits sizable
fluctuations at temperatures below the scale set by the gap. By contrast, open
geometries allow for sizable thermal fluctuations at any value of the
temperature. 

Concentrating on the inclined-plates geometry in $D$ dimensions, the
temperature dependence {of the Casimir force} can become stronger by one
power in the temperature parameter (implying thermal corrections which can be
an order of magnitude larger than for a closed geometry). The inclined-plates
geometry is particularly interesting as the limit of a semi-infinite plate
parallel to an infinite plate (1si configuration) is somewhat in-between open
and closed geometries: the open part of the spectrum only arises due to the
edge of the semi-infinite plate. Interestingly, the resulting thermal
correction numerically shows a power-law temperature dependence with a
non-integer exponent $\sim T^{D-0.3}$.

{The strongest temperature dependence $\sim T^{D-2}$ in the
  low-temperature limit occurs for the Casimir torque of the inclined-plates
  configuration. This is, because it arises from the leading thermal
  correction of the interaction energy which contributes to the torque but not
  to the Casimir force. }

Our results have been derived for the case of a fluctuating scalar field
obeying Dirichlet boundary conditions on the surfaces. Whereas this model system
should not be considered as a quantitatively appropriate model for the real
electromagnetic Casimir effect, our general conclusions about the
geometry-temperature interplay are not restricted to the Dirichlet scalar
case. On the contrary, all our arguments based on the presence or absence of a
spectral gap will also be valid for the electromagnetic case. Whether or not
the case of Neumann or electromagnetic boundary conditions leads to different
power-law exponents for the temperature dependence of the ``geothermal''
phenomena remains an interesting question for future research. 

In view of the fact that most (strictly speaking all) experiments are
performed in open geometries, e.g., the sphere-plate geometry, at room
temperature, an analysis of the geometry-temperature interplay of these 
experimentally relevant configurations is most pressing. 

\appendix

\section{Inclined plates, $\varphi\rightarrow 0$ limit at finite temperature}

We have analyzed the $\varphi\rightarrow 0$ behavior of inclined plates at
zero temperature in Sect. \ref{sec:incl-plat-varph}. Here, we consider the
same limit for the thermal correction to the energy. The decomposition of the
1si Casimir energy into bulk and edge contributions can also be performed for
the thermal corrections,
$$
\Delta E_\mathrm{c}^\mathrm{1si,\varphi}(\xi)=\Delta
E_\mathrm{c}^\mathrm{edge,\varphi}(\xi)+ \Delta
E_\mathrm{c}^\mathrm{\parallel,\varphi}(\xi), 
$$
where 
\begin{widetext}
\begin{align}\label{wpir-iplaft-f1}
\Delta E_\mathrm{c}^{\parallel,\varphi}(\xi) =& -\frac{L_{y}\csc(\varphi)}{(4
  \pi )^{D/2} a^{D-2}} 
\left\langle\left[-\Gamma \left(\frac{D-2}{2}\right) \zeta (D-2)(2 \xi)^{D-2}
  \lambda_{x}\bgam_{x_\mathrm{min}}(\varphi) -\zeta (D-1) \Gamma
  \left(\frac{D-1}{2}\right) (2 \xi)^{D-1}\sqrt{\pi} \right.\right.\notag 
\\
&\left.\left.-\sum_{n=1}^{\infty}\int_{\bgam_{x_\mathrm{min}}}^{\bgam_{x_\mathrm{max}}}
\!\!\!\!\!\! \mathrm{d} x\,\,\Big\{\bgam_\mathrm{m}^{D-2}(x)E_{2-\frac{D}{2}}\left(\frac{n^2
  \bgam_\mathrm{m}^2 (x)}{4 \xi^2}\right) (x \cos
(\varphi)-\bgam_{x_\mathrm{min}}(\varphi )) - \bgam_\mathrm{m}^{D-1}(x)
E_{\frac{3-D}{2}}\left(\frac{n^2 \bgam_\mathrm{m}^2(x)}{4 \xi^2}\right) \Big\}
\right]\right\rangle, 
\end{align}
%
%
with $\bgam_\mathrm{m}(x)= x \cos (\varphi )+\sin
(\varphi ) \bgam_{z_{\mathrm{max}}}(x)-\bgam_{x_\mathrm{min}}(\varphi)$ and
%
 %
\begin{align}\label{wpir-iplaft-f2}
\Delta E_\mathrm{c}^{\mathrm{edge},\varphi}(\xi) = -\frac{L_{y}}{(4 \pi
  )^{D/2} a^{D-2}}&\left[ \Gamma \left(\frac{D-2}{2}\right) \zeta (D-2)
  (2\xi)^{D-2} \left\langle
  \int_{\bgam_{x_\mathrm{min}}}^{\bgam_{x_\mathrm{max}}}\!\!\!\!\!\!
  \mathrm{d} x\, 
  \bgam_{z_{\mathrm{max}}}(x) \right\rangle\right.\notag 
\\
&\left.- \sum_{n=1}^{\infty}\left\langle 
\int_{\bgam_{x_\mathrm{min}}}^{\bgam_{x_\mathrm{max}}}\!\!\!\!\!\!\!
 \mathrm{d}x \,\bgam_\mathrm{m}^{D-2}(x)
E_{2-\frac{D}{2}}\left(\frac{n^2\bgam_\mathrm{m}^2(x)}{4 \xi^2}\right)
\bgam_{z_{\mathrm{max}}}(x)\right\rangle\right].
\end{align}
\end{widetext}
Whereas $\Delta E_\mathrm{c}^\mathrm{edge,\varphi}(\xi)$ remains finite,
$\Delta E_\mathrm{c}^\mathrm{\parallel,\varphi}(\xi)$ shows a divergent
behavior as $\varphi\rightarrow 0$. Let us therefore concentrate on $\Delta
E_\mathrm{c}^\mathrm{\parallel,\varphi}(\xi)$, in order to isolate the source
of the apparent divergence which is related to the order of limits of
$L_z\rightarrow \infty$ and $\varphi \rightarrow 0$. In the case of 
inclined plates, $(L_z\varphi)$ is infinite for all $\varphi\neq 0$, resulting
in a $1/\varphi$ divergent energy density (energy per length) for
$\varphi\rightarrow 0$. Parallel plates, on the other hand have a finite
energy density (energy per area) and $(L_z\varphi)=0$.

\begin{figure}[t]
\begin{center}
\scalebox{0.7}{
\begin{picture}(0,0)%
\includegraphics{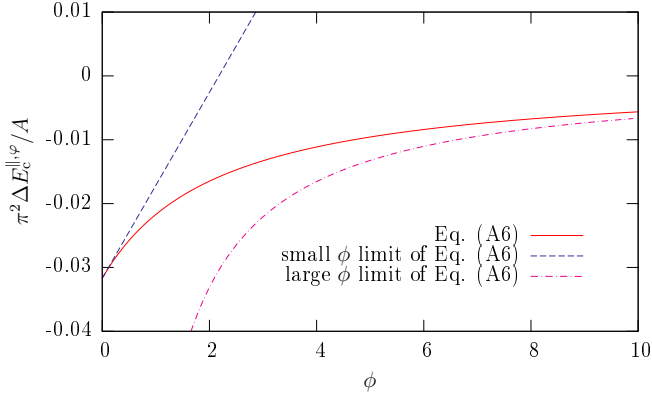}%
\end{picture}%
\begingroup
\setlength{\unitlength}{0.0200bp}%
\begin{picture}(18000,10800)(0,0)%
\large
\put(2475,1650){\makebox(0,0)[r]{\strut{}-0.04}}%
\put(2475,3370){\makebox(0,0)[r]{\strut{}-0.03}}%
\put(2475,5090){\makebox(0,0)[r]{\strut{}-0.02}}%
\put(2475,6810){\makebox(0,0)[r]{\strut{}-0.01}}%
\put(2475,8530){\makebox(0,0)[r]{\strut{} 0}}%
\put(2475,10250){\makebox(0,0)[r]{\strut{} 0.01}}%
\put(2750,1100){\makebox(0,0){\strut{} 0}}%
\put(5635,1100){\makebox(0,0){\strut{} 2}}%
\put(8520,1100){\makebox(0,0){\strut{} 4}}%
\put(11405,1100){\makebox(0,0){\strut{} 6}}%
\put(14290,1100){\makebox(0,0){\strut{} 8}}%
\put(17175,1100){\makebox(0,0){\strut{} 10}}%
\put(550,5950){\rotatebox{90}{\makebox(0,0){\strut{}$\pi^2 \Delta  E_\mathrm{c}^{\parallel,\varphi}/A$}}}%
\put(9962,275){\makebox(0,0){\strut{}$\phi$}}%
\put(14015,4230){\makebox(0,0)[r]{\strut{} Eq. (\ref{wpir-iplaft-f5})}}%
\put(14015,3680){\makebox(0,0)[r]{\strut{}small $\phi$ limit of  Eq. (\ref{wpir-iplaft-f5})}}%
\put(14015,3130){\makebox(0,0)[r]{\strut{}large $\phi$ limit of Eq. (\ref{wpir-iplaft-f5})}}%
\normalsize
\end{picture}%
\endgroup
}
\end{center}
\caption{Qualitative behavior of the thermal Casimir contribution of the bulk
  in the inclined-plates case Eq.~(\ref{wpir-iplaft-f6}) (red line) and its
  small-$\phi$ (dashed blue line) and large-$\phi$ (dot-dashed magenta line)
  limit, respectively. Note the divergent
  $1/\phi$ behavior of the large-$\phi$ limit which corresponds to the
  divergent (as $\varphi\rightarrow 0$) energy per edge length in the
  inclined-plates formulas.  For this illustration, we have chosen
  $D=4,\ a=\beta=\lambda_{x}=1$, ignoring the worldline average in
  Eq.~(\ref{wpir-iplaft-f6}) for simplicity.}\label{Fig-incLimit-3}
\end{figure}

In the following, we show how to obtain an analytic transition from
$(L_z\varphi)\rightarrow \infty$ to $(L_z\varphi)\rightarrow 0$ for small
$\varphi$ by working with large but finite $L_z$, and taking
$L_z\rightarrow\infty$ at the end of the calculation.  The first limit results
in a divergent energy density per unit edge length of the inclined plates as
$\varphi\rightarrow 0$,
\begin{widetext}
\begin{align}\label{wpir-iplaft-f3}
\Delta E_\mathrm{c}^{\parallel,\varphi\rightarrow 0}(\xi) &=- \frac{L_{y}}{2
  a^{D-2} \varphi  (4 \pi )^{D/2}} \Big[\langle \lambda^2\rangle \Gamma
  \left(\frac{D-2}{2}\right) \zeta (D-2) (2 \xi)^{D-2}-2 (2 \xi)^{D-1} \zeta
  (D-1) \Gamma \left(\frac{D-1}{2}\right) \sqrt{\pi}
\\
& +\Gamma \left(\frac{D}{2}\right) \zeta (D) (2 \xi)^D-\left\langle
\lambda^D\sum_{n=1}^{\infty}\left(E_{1-\frac{D}{2}}\left(\frac{\lambda^2
  n^2}{4 \xi^2}\right)+E_{2-\frac{D}{2}}\left(\frac{\lambda^2 n^2}{4
  \xi^2}\right)-2 E_{\frac{3}{2}-\frac{D}{2}}\left(\frac{\lambda^2 n^2}{4
  \xi^2}\right)\right)\right\rangle \Big]+\mathcal{O}\left(1\right). \notag 
\end{align}
\end{widetext}
The second limit corresponds to the finite energy density of exact parallel
plates (\ref{FT-pp-1}).

In Eqs.~(\ref{wpir-iplaft-f1}-\ref{wpir-iplaft-f3}), the $z$ integration was
performed first. Let us now do the proper time integration first. The $\theta$
function~(\ref{T0-ip-l-3}), valid for small $\varphi$, reflects itself in the
lower end of the proper time integral:
\begin{align}\label{wpir-iplaft-f4}
\Delta E_\mathrm{c}^{\parallel,\varphi\rightarrow 0}&(a,\beta) =
-\frac{L_{y}}{(4 \pi )^{D/2}}\left\langle\sum_{n=1}^{\infty}\right.\notag 
\\
 &\times\int_1^{\infty } \frac{\exp\left(-\frac{\beta ^2 x^2 n^2}{4 T (a+z
    \varphi)^2}\right)\mathrm{d}\mathcal{T}}{\mathcal{T}^{\frac{D+1}{2}}}\notag 
\\
 & \times\int_0^{L_z/2}\left.
   \frac{\mathrm{d}z}{(a+z \varphi )^{D-1}}
   \int_{0}^{\lambda_{x}}x^{D-1} \mathrm{d}x\Big]\right\rangle. 
\end{align}
The proper time integration yields for Re$[D]>1$:
\begin{align}\label{wpir-iplaft-f4b}
&\int_1^{\infty } \frac{\exp\left(-\frac{\beta ^2 x^2 n^2}{4 \mathcal{T} (a+z
      \varphi)^2}\right)\mathrm{d}\mathcal{T}}{\mathcal{T}^{\frac{D+1}{2}}} =
\\
&\left(\frac{ 2(a+z \varphi)}{x n \beta  }\right)^{D-1} \Gamma \left(\frac{D-1}{2}\right)-E_{\frac{3-D}{2}}\left(\frac{x^2 \beta
   ^2 n^2}{4 (a+z \varphi )^2}\right).  \notag 
\end{align}
Inserting Eq.~(\ref{wpir-iplaft-f4b}) into  Eq.~(\ref{wpir-iplaft-f4}) leads to
\begin{widetext}
\begin{align}\label{wpir-iplaft-f5}
\frac{\pi ^{\frac{D}{2}} \Delta E_\mathrm{c}^{\parallel,\varphi\rightarrow
    0}(a,\beta)}{A} &=\sum_{n=1}^{\infty}\left\langle\frac{a^2 \left(\Gamma
  \left(\frac{D}{2},\frac{\lambda_{x}^2 n^2 \beta ^2}{4
    a^2}\right)-\Gamma \left(\frac{D}{2},\frac{\lambda_{x}^2 n^2 \beta
    ^2}{(2 a+\phi )^2}\right)\right)}{\phi (n \beta
  )^{D}}+\frac{\lambda_{x}^2 \left(\Gamma
  \left(\frac{D-2}{2},\frac{\lambda_{x}^2 n^2 \beta ^2}{4
    a^2}\right)-\Gamma \left(\frac{D-2}{2},\frac{\lambda_{x}^2 n^2
    \beta ^2}{(2 a+ \phi )^2}\right)\right)}{4 \phi (n \beta )^{D-2}}\right.\notag
\\ &-\frac{a \lambda_{x} \left(\Gamma
  \left(\frac{D-1}{2},\frac{\lambda_{x}^2 n^2 \beta ^2}{4
    a^2}\right)-\Gamma \left(\frac{D-1}{2},\frac{\lambda_{x}^2 n^2
    \beta ^2}{(2 a+\phi )^2}\right)\right)}{\phi (n\beta )^{D-1}}+\frac{\phi
  \left(\Gamma \left(\frac{D}{2}\right)-\Gamma
  \left(\frac{D}{2},\frac{\lambda_{x} ^2 n^2 \beta ^2}{(2 a+\phi
    )^2}\right)\right)}{4 (n \beta)^{D}} \notag 
\\ &\left.+\frac{ \lambda_{x}
  \Gamma \left(\frac{D-1}{2},\frac{\lambda_{x}^2 n^2 \beta ^2}{(2
    a+\phi )^2}\right)}{2 (n \beta )^{D-1}} - \frac{a \Gamma
  \left(\frac{D}{2},\frac{L^2 n^2 \beta ^2}{(2 a+ \phi )^2}\right)}{ (n \beta
  )^{D}} \right\rangle-\frac{\sqrt{\pi} \Gamma
  \left(\frac{D-1}{2}\right)\zeta(D-1)}{2 \beta ^{D-1}} + \frac{a\Gamma
  \left(\frac{D}{2}\right)\zeta(D)}{ \beta^{D}},
\end{align}
\end{widetext}
where $\phi\equiv L_z\varphi$ and $A=L_{y}L_z/2$. One can show that the first
three terms of Eq.~(\ref{wpir-iplaft-f5}) are of order $\mathcal
O(\phi^2)$. The forth term is clearly $\mathcal O(\phi)$. The last line can be
converted into the parallel-plates energy density (\ref{FT-pp-1}) by
neglecting $\phi$ with respect to $a$ and using the identity $z^\alpha
E_{1-\alpha}(z)=\Gamma (\alpha,z)$; the error is of order $\mathcal
O(\phi^2)$. The first-order correction to the parallel-plates case is
therefore encoded in the fourth term. The second-order correction is in the
first three terms since the $\phi^2$ terms cancel each other in the
remainder. In this limit ($|\phi|\ll 1$), all sums converge for
$\mathrm{Re}[D]>2$.

Let us rearrange (\ref{wpir-iplaft-f5}) so as to investigate the
$\phi\rightarrow \infty$ case with $\varphi$ being small but finite:
\begin{widetext}
\begin{align}\label{wpir-iplaft-f6}
\frac{\pi ^{\frac{D}{2}} \Delta  E_\mathrm{c}^{\parallel,\varphi}(a,\beta)}{A}
&=\sum_{n=1}^{\infty}\left\langle\frac{(4 a+\phi ) \left(\Gamma
  \left(\frac{D}{2}\right)-\Gamma \left(\frac{D}{2},\frac{\lambda_{x}^2
    n^2 \beta ^2}{(2 a+\phi)^2}\right)\right)}{4 (n\beta
  )^{D}}+\frac{\lambda_{x}\left(\Gamma
  \left(\frac{D-1}{2},\frac{\lambda_{x}^2 n^2 \beta ^2}{(2 a+\phi
    )^2}\right)-\Gamma
  \left(\frac{D-1}{2}\right)\right)}{2(n\beta)^{D-1}}\right.\notag 
\\
&-\frac{\lambda_{x}^2 \Gamma \left(\frac{D-2}{2},\frac{\lambda_{x}^2 n^2 \beta^2}{(2 a+\phi )^2}\right) }{4 \phi (n \beta )^{D-2}}
+\frac{a \lambda_{x} \Gamma \left(\frac{D-1}{2},\frac{\lambda_{x}^2 n^2 \beta ^2}{(2 a+\phi )^2}\right) }{\phi (n \beta )^{D-1}}
-\frac{a^2 \Gamma \left(\frac{D}{2},\frac{\lambda_{x}^2 n^2 \beta ^2}{(2 a+\phi )^2}\right) }{\phi (n \beta )^{D}}\notag
\\
&\left.+\frac{\lambda_{x}^2 \Gamma
   \left(\frac{D-2}{2},\frac{\lambda_{x}^2 n^2 \beta ^2}{4 a^2}\right) }{4 \phi (n \beta )^{D-2}}-\frac{a \lambda_{x} \Gamma
   \left(\frac{D-1}{2},\frac{\lambda_{x}^2 n^2 \beta ^2}{4 a^2}\right) }{\phi
  (n \beta )^{D-1}}+\frac{a^2 \Gamma \left(\frac{D}{2},\frac{\lambda_{x}^2 n^2
    \beta ^2}{4 a^2}\right) }{\phi (n \beta)^{D}}\right\rangle. 
\end{align}
\end{widetext}
The large-$\phi$ behavior of the first two terms can be obtained through
Poisson summation\footnote{The large-$x$ limit of $\sum_{n=1}^\infty f(n/x)$
  yields $(-f(0)+\sqrt{2\pi} \hat{f}(0)x)/2$ where $\hat{f}$ is the Fourier
  transform of $f$.} and reads
\begin{align}\label{wpir-iplaft-f7}
\frac{(4 a+\phi) (c_1+c_2 (2 a+\phi ) )}{(2a+\phi )^D} + \frac{(c_3+c_4 (2 a+\phi ) )}{(2a+\phi )^{D-1}},
\end{align}
where $c_1,\dots, c_4$ are constants, the values of which are of no
importance. We see that Eq.~(\ref{wpir-iplaft-f7}) vanishes for $D>2$. For $D>3$
the terms vanish even if multiplied by the infinite length $L_z$. Remember
that the inclined-plates formulae at finite temperature are valid for $D>3$
as well.

In order to keep the remaining terms of Eq.~(\ref{wpir-iplaft-f6}) finite, we
multiply both sides with the infinite length $L_z$ converting the vanishing
Casimir energy per area into the finite energy per length. The Poisson
summation of the second line of Eq.~(\ref{wpir-iplaft-f6}) results in
\begin{widetext}
\begin{align}\label{wpir-iplaft-f8}
\lim\limits_{\phi \to \infty }\sum_{n=1}^{\infty}&\left(-\frac{\lambda_{x}^2
  \Gamma \left(\frac{D-2}{2},\frac{\lambda_{x}^2 n^2 \beta^2}{(2 a+\phi
    )^2}\right) }{4 \varphi (n \beta )^{D-2}} 
+\frac{a \lambda_{x} \Gamma \left(\frac{D-1}{2},\frac{\lambda_{x}^2 n^2 \beta
    ^2}{(2 a+\phi )^2}\right) }{\varphi (n \beta )^{D-1}} 
-\frac{a^2 \Gamma \left(\frac{D}{2},\frac{\lambda_{x}^2 n^2 \beta ^2}{(2 a+\phi )^2}\right) }{\varphi (n \beta )^{D}}\right)=\notag
-\frac{\lambda_{x}^2 \Gamma \left(\frac{D-2}{2}\right) \zeta(D-2)}{4 \varphi
  \beta^{D-2}}
\\
&
+\frac{a \lambda_{x} \Gamma \left(\frac{D-1}{2}\right) \zeta(D-1)}{\varphi \beta^{D-1}}
-\frac{a^2 \Gamma \left(\frac{D}{2}\right) \zeta(D)}{\varphi \beta^{D}}
+\frac{d_1+d_2 (2 a+\phi )}{\varphi(2 a+\phi )^{D-2}}+\frac{d_3+d_4 (2 a+\phi )}{\varphi(2 a+\phi )^{D-1}}+\frac{d_5+d_6 (2 a+\phi )}{\varphi(2 a+\phi )^{D}},
\end{align}
\end{widetext}
where $d_1,\dots,d_6$ are constants. These terms containing $d_i$'s vanish for
$D>3$ and $\phi\rightarrow \infty$. Applying the identity $z^a
E_{1-a}(z)=\Gamma (a,z)$ to the last three terms in
Eq.~(\ref{wpir-iplaft-f6}), we rediscover the inclined-plates formula
(\ref{wpir-iplaft-f3}) from Eqs.~(\ref{wpir-iplaft-f6}),
(\ref{wpir-iplaft-f8}) valid for small angles $\varphi$. From
Eqs.~(\ref{wpir-iplaft-f7}) and (\ref{wpir-iplaft-f8}), one can infer that the
first correction to Eq.~(\ref{wpir-iplaft-f3}) is of order $O(1/\phi^{D-3})$.

\acknowledgments

The authors are grateful to Klaus Klingm\"uller for interesting discussions. AW
acknowledges support by the Landesgraduiertenf\"orderung Baden-W\"urttemberg,
by the Heidelberg Graduate School of Fundamental Physics, and by the DFG under
contract Gi 328/3-2. HG was supported by the DFG under contract No. Gi 328/1-4
(Emmy-Noether program), Gi 328/5-1 (Heisenberg program).

\end{document}